\documentclass[fleqn,12pt]{article}
\usepackage{hyperref,rotating,multicol,bm}

\usepackage{amsmath, amssymb, multirow, graphicx,footnote,caption,color}

\makeatletter
\newcommand{\Rmnum}[1]{\expandafter\@slowromancap\romannumeral #1@}
\makeatother
\begin{document}\sloppy
\title{\bf{A Simplified Stochastic EM Algorithm for Cure Rate Model with Negative Binomial Competing Risks: An Application to Breast Cancer Data}}
\author{\bf{Suvra Pal}\\\\
Department of Mathematics, University of Texas at Arlington, \\
Arlington, Texas 76019, United States}
\date{}

\maketitle

\begin{abstract}

\noindent In this paper, a long-term survival model under competing risks is considered. The unobserved number of competing risks is assumed to follow a negative binomial distribution that can capture both over- and under-dispersion. Considering the latent competing risks as missing data, a variation of the well-known expectation maximization (EM) algorithm, called the stochastic EM algorithm (SEM), is developed. It is shown that the SEM algorithm avoids calculation of complicated expectations, which is a major advantage of the SEM algorithm over the EM algorithm. The proposed procedure also allows the objective function to be split into two simpler functions, one corresponding to the parameters associated with the cure rate and the other corresponding to the parameters associated with the progression times. The advantage of this approach is that each simple function, with lower parameter dimension, can be maximized independently. An extensive Monte Carlo simulation study is carried out to compare the performances of the SEM and EM algorithms. Finally, a breast cancer survival data is analyzed and it is shown that the SEM algorithm performs better than the EM algorithm. \\


\end{abstract}

\section{Introduction}

Advancements in the treatments of certain types of diseases (e.g., cancer, heart disease, etc.) have led to a noteworthy number of patients to respond favorably to the treatment, thereby not showing recurrence until the end of a long follow-up time. These patients are termed as recurrence-free survivors. It is quite possible that some of these patients will not show recurrence for a decently long period after the follow-up time since they may have reached a stage where the disease is undetectable and harmless. These patients, among the recurrence-free survivors, are termed as long-term survivors or ``cured''. The estimation of the proportion of cured patients cannot be readily obtained from a given lifetime data since it is not possible to identify which of the recurrence-free survivors can be considered as long-term survivors. This is because a patient who is susceptible to disease recurrence may show no recurrence until the end of the follow-up time. The estimation of a treatment-specific cured proportion (or cure rate), however, is important to see the trend in the survival of patients suffering from a particular disease. It is also an important measure to judge the treatment's efficacy and its adoption in practice, as opposed to the standard treatment.

The first studied cure rate model dates back to mid 90s,$^{1,2}$ which is known in the literature as the mixture cure rate model. According to the mixture cure rate model, the population survival function of a time-to-event variable $Y$ is given by 
\begin{equation}
S_p(y) = p_0 + (1-p_0)S_s(y),
\label{mix}
\end{equation}
where $p_0$ is the cure rate and $S_s(\cdot)$ is the survival function of the susceptible patients only, which is a proper survival function. To incorporate a competing risks scenario, where several risk factors can compete to produce the event of interest, researchers have proposed the promotion time cure rate model.$^3$ In this case, the population survival function is given by 
\begin{equation}
S_p(y) = e^{-\eta(1-S(y))},
\label{prom}
\end{equation}
where $\eta$ represents the mean number of competing risks and $S(\cdot)$ represents the common survival function of the promotion times corresponding to each risk factor. The cure rate, in this case, is given by $e^{-\eta}$. To study the effect of risk factors or covariates on the cure rate, one can relate $p_0$ in \eqref{mix} to a set of risk factors using a logistic link function.$^{4,5}$ Similarly, for the promotion time cure rate model, one can relate $\eta$ to a set of risk factors using a log-linear link function.$^6$

There is a class of cure rate models which include the mixture and promotion time models as special cases.$^7$ Several approaches have been proposed in the literature to develop the associated inference for some flexible cure rate models that include the aforementioned mixture and promotion time cure rate models. In this regard, interested readers may refer to parametric,$^{8,9}$ semi-parametric$^{10-12}$ and non-parametric approaches.$^{13}$ With regard to elimination of tumor cells after a prolonged treatment, there are specific references that mention about the radio-biological evidence on the temporal characteristics of enzymatic repair.$^{14}$ Motivated by this elimination process, researchers described the biological process of elimination of tumor cells after some specific treatment and proposed the destructive weighted Poisson cure rate models.$^{15-20}$ Work has also been done on a new Bayesian flexible cure rate survival model that uses Markov Chain Monte Carlo methods to develop Bayesian inference.$^{21}$

In a competing risks scenario, also called a competing causes scenario, both the number of competing risks and the lifetime, also called the progression time, associated with each competing risk are unobserved, that is, they are considered as latent variables. So, both these latent variables need to be suitably modeled. Let $M$ denote the number of risk factors competing to give rise to an event of interest, for instance, death due to a particular cancer, recurrence of a tumor, etc. Then, condition on $M=m$, we further denote $W_j$ to be the lifetime due to the $j-$th risk factor, $j=1,\ldots,m$. Moreover, we also assume that these $W_j$'s are distributed independent of $M$. Under this setting, we only observe the minimum of all lifetimes, that is, we observe
\begin{equation} \label{Ymin}
Y=\min\{W_0,W_1,\ldots,W_M\},
\end{equation}
where $W_0$ is an infinite lifetime corresponding to no competing risks, i.e., $M=0.$ Such a lifetime $W_0$ is defined as $P[W_0 = \infty] = 1,$ which captures the cure proportion. We can then simply define the cure proportion or the cure rate as $p_0=P(M=0)$. Developing a cure rate model then includes modeling the number of competing risks, the lifetime distribution under each competing risk, the distribution of $Y$, and, finally, the cure rate $p_0$. As far as developing the likelihood inference and finding the maximum likelihood estimates (MLEs) are concerned in the given context, the expectation maximization (EM) algorithm is a very popular technique.$^{22-25}$

In this paper, we consider a competing risks scenario and assume the number of competing risks to follow a negative binomial distribution. In this way, we can handle both over- and under-dispersion that we usually encounter when modeling count data on competing risks.$^{26}$ Instead of assuming the lifetime distributions $W_j, j = 1,2,\cdots,M,$ of competing risks to be identical or homogeneous, as commonly assumed in the literature,$^{6,22,23}$ we let the lifetime distribution to depend on a set of covariates. This not only allows us to study the effect of covariates on the lifetime distribution but also allows us to formally test for the suitability of the homogeneous lifetime distribution assumption. For the estimation of the model parameters, we follow a simplified likelihood approach.$^{27}$ In this approach, two things are noteworthy. First, the missing data is introduced through the latent competing risks variable, which is a completely different approach from the traditional approach,$^{4,9}$ where the missing data is introduced through the latent cured status variable. Second, the objective function to be maximized can be easily split into two simpler functions, one corresponding to the parameters associated with the cure rate and the other corresponding to the parameters associated with the competing risk lifetime distribution. This allows each of these two functions to be maximized independently, which is more convenient and preferable than maximizing one complicated function with all model parameters. Motivated by the presence of missing data, we propose a variation of the well studied expectation maximization (EM) algorithm,$^{6,9}$ called the stochastic EM (SEM) algorithm. This variation comes with a major advantage; it avoids calculation of complicated conditional expectations and it is enough to calculate the conditional distribution of the missing data.$^{28}$ 

The rest of this paper is organized as follows. In Section 2, we describe the cure rate model with negative binomial competing risks and non-homogeneous lifetime distribution. In Section 3, we define the observed and complete likelihood functions, which are required for the development of the SEM algorithm. In Section 4, we develop the steps of the proposed SEM algorithm in detail and point out the advantages of the SEM algorithm over the EM algorithm. In Section 5, we present the results of an extensive Monte Carlo simulation study where we compare the performances of the SEM and EM algorithms. In Section 6, we illustrate the proposed method using a breast cancer survival data and show that the SEM algorithm performs better than the EM algorithm. Finally, in Section 7, we make some concluding remarks and discuss some potential future research problems.

\section{Cure rate model with negative binomial competing risks}

Let us assume the unobserved number of competing risks $M$ to follow a negative binomial distribution with probability mass function (pmf) 
\begin{equation}
p_m = P[M=m;\eta,\phi]=\frac{\Gamma(m+\frac{1}{\phi})}{\Gamma(\frac{1}{\phi})m!}\bigg(\frac{\phi\eta}{1+\phi\eta}\bigg)^m \bigg(\frac{1}{1+\phi\eta}\bigg)^{\frac{1}{\phi}}, \ \  m=0,1,2,\ldots,
\label{nb}
\end{equation}
where $\eta > 0,$ $\phi > -\frac{1}{\eta}$, and $\Gamma(p) = \int_0^\infty e^{-x}x^{p-1}dx$ is the complete gamma function. We call \eqref{nb} as a negative binomial distribution with parameters $r=\frac{1}{\phi}$ and $p=\frac{1}{1+\phi\eta}$. From \eqref{nb}, it is easy to see that $E(M) = \eta$ and $V(M) = \eta + \phi \eta^2.$ Hence, values of $\phi > 0$ correspond to over-dispersion, whereas values of $\phi < 0$ correspond to under-dispersion, both relative to the Poisson distribution. When $\phi\rightarrow 0,$ the pmf in \eqref{nb} approaches to that of a Poisson random variable with mean $\eta.$ For $\phi=1,$ the pmf in \eqref{nb} reduces to that of a geometric distributed random variable with parameter $\frac{1}{1+\eta}$.$^{29		}$

We assume the lifetime $W_j$ due to each competing risk to follow a parametric distribution, although one can assume some non-parametric or semi-parametric models for $W_j$ as well. Next, we choose a suitable parameter of the chosen lifetime distribution and associate it with a set of covariates $\boldsymbol x$ using an appropriate link function $g(\boldsymbol x^\prime\boldsymbol\alpha),$ where $\boldsymbol\alpha$ denotes the vector of regression coefficients. Let us denote the other parameters of the chosen lifetime distribution by $\boldsymbol\gamma$. Notationally, let $S(w;\boldsymbol x)$, $f(w;\boldsymbol x)$ and $F(w;\boldsymbol x)$ denote the common survival function, density function and cumulative distribution function, respectively, of $W_j, j = 1,2,\cdots,M$. Note that although we assume identical lifetime distributions for all competing risks corresponding to a given subject, the lifetime distributions across susceptible patients are non-identical. This captures the non-homogeneity in patient population who are susceptible to the event of interest. Under this competing risks scenario, the survival function of the time-to-event variable $Y$ in \eqref{Ymin}, also called the long-term survival function, is given by
\begin{equation}
S_p(y) = S_p(y;\boldsymbol x) = \sum_{m=0}^\infty \{S(y;\boldsymbol x) \}^m p_m, 
\end{equation}
which is the probability generating function (pgf) of $M$, evaluated at $S(y;\boldsymbol x)$. Now, using \eqref{nb}, it can be easily shown that 
\begin{eqnarray}
S_p(y;\boldsymbol x) = \bigg\{\frac{1}{1+ \phi\eta F(y;\boldsymbol x)} \bigg\}^{\frac{1}{\phi}}.
\label{Sp}
\end{eqnarray}
Note that $S_p(y;\boldsymbol x)$ is not a proper survival function since $\lim_{y\rightarrow\infty} S_p(y;\boldsymbol x) > 0$, given that $F(y;\boldsymbol x)$ is a proper distribution function. The long-term density function, or, equivalently, the population density function can be calculated as
\begin{eqnarray}
f_p(y) = f_p(y;\boldsymbol x) = - S_p^\prime(y;\boldsymbol x) =  \eta f(y;\boldsymbol x) \bigg\{ \frac{1}{1+\phi\eta F(y;\boldsymbol x)} \bigg\}^{\frac{1}{\phi}+1}.
\label{fp}
\end{eqnarray}
From \eqref{Sp}, the cure rate or the long-term survival probability is given by
\begin{eqnarray}
p_0 = \lim_{y\rightarrow\infty} S_p(y;\boldsymbol x) =  \bigg\{\frac{1}{1+ \phi\eta} \bigg\}^{\frac{1}{\phi}}.
\label{p0}
\end{eqnarray}
Note that the parameters in the negative binomial cure rate model in \eqref{Sp} carry biological interpretations. The parameter $\eta$ represents the mean number of competing risks, whereas the parameter $\phi$ accounts for the inter-individual variance of the number of competing risks. Now, $\eta$ being the mean number of competing risks, the cure rate should decrease with an increase in the number of competing risks. This behavior is captured in \eqref{p0} since $p_0$ is indeed a decreasing function of $\eta$. In a practical scenario, the cure rate $p_0$ should also depend on a set of covariates. For this purpose, and to study the effect of covariates on the cure rate, we link the parameter $\eta$ with another set of covariates $\boldsymbol z$ using the following function
\begin{eqnarray}
\eta = \eta(\boldsymbol z) = \exp(\boldsymbol z^\prime\boldsymbol\beta),
\label{eta}
\end{eqnarray}
where $\boldsymbol\beta$ represents the vector of regression coefficients. From \eqref{eta}, we can rewrite the expressions as
\begin{eqnarray}
p_0 =p_0(\boldsymbol z) =  \bigg\{\frac{1}{1+ \phi\eta(\boldsymbol z)} \bigg\}^{\frac{1}{\phi}},
\end{eqnarray}
\begin{eqnarray}
S_p(y) = S_p(y;\boldsymbol x,\boldsymbol z) = \bigg\{\frac{1}{1+ \phi\eta (\boldsymbol z) F(y;\boldsymbol x)} \bigg\}^{\frac{1}{\phi}},
\label{Sp2}
\end{eqnarray}
and
\begin{eqnarray}
f_p(y) = f_p(y;\boldsymbol x,\boldsymbol z) = \eta (\boldsymbol z) f(y;\boldsymbol x) \bigg\{ \frac{1}{1+\phi\eta (\boldsymbol z) F(y;\boldsymbol x)} \bigg\}^{\frac{1}{\phi}+1}.
\label{fp2}
\end{eqnarray}
Note that $\boldsymbol x$ and $\boldsymbol z$ may share common elements. Let us denote the vector of unknown parameters by $\boldsymbol\theta = (\phi,\boldsymbol\alpha^\prime,\boldsymbol\beta^\prime,\boldsymbol\gamma^\prime)^\prime$.

\section{Observed and complete likelihood functions}

We consider a practical scenario where the lifetime $Y$ in \eqref{Ymin} may not be completely observed and is subject to right censoring. If we denote by $C$ the censoring time, then, the observed lifetime is defined as $T = \min\{Y,C \}$. Furthermore, we can define the right censoring indicator as $\delta = I[Y < C]$, where $I[A]$ takes the value 1 if the event $A$ occurs, and is 0 otherwise. Thus, if we have $n$ patients in a study, the observed data can be defined as $\boldsymbol D_o = \{(t_i,\delta_i,\boldsymbol x_i,\boldsymbol z_i), i = 1,2,\cdots,n\}$. Based on $\boldsymbol D_o$, the observed data likelihood function can be written as
\begin{eqnarray*}
L_o(\boldsymbol\theta| \boldsymbol D_o) = \prod_{i=1}^n \{f_p(t_i;\boldsymbol x_i,\boldsymbol z_i)\}^{\delta_i} \{S_p(t_i;\boldsymbol x_i,\boldsymbol z_i)\}^{1-\delta_i}.
\end{eqnarray*}
Hence, the observed data log-likelihood function can be written as
\begin{eqnarray*}
l_o(\boldsymbol\theta| \boldsymbol D_o) &=& \sum_{i=1}^n \{ \delta_i \log f_p(t_i;\boldsymbol x_i,\boldsymbol z_i) + ( 1-\delta_i) \log S_p(t_i;\boldsymbol x_i,\boldsymbol z_i) \}\\
&=& \sum_{i=1}^n \bigg[ \delta_i\bigg\{  \log \eta(\boldsymbol z_i) + \log f(t_i;\boldsymbol x_i) + \bigg( \frac{1}{\phi}+1 \bigg)\log\bigg(\frac{1}{1+\phi \eta(\boldsymbol z_i)F(t_i;\boldsymbol x_i)} \bigg) \bigg\} \\
&&+ (1-\delta_i) \frac{1}{\phi} \log\bigg( \frac{1}{1+\phi\eta(\boldsymbol z_i)F(t_i;\boldsymbol x_i)} \bigg)   \bigg]\\
&=& \sum_{i=1}^n \bigg[ \delta_i\{  \log\eta(\boldsymbol z_i) + \log f(t_i;\boldsymbol x_i) - \log(1+ \phi\eta(\boldsymbol z_i)F(t_i;\boldsymbol x_i)) \} - \frac{1}{\phi} \log(1+ \phi\eta(\boldsymbol z_i)F(t_i;\boldsymbol x_i))   \bigg].
\end{eqnarray*}
To estimate the model parameters, one can directly maximize the above log-likelihood function using any readily available software packages in R, such as ``nlm()" or ``optim()", among others. However, these readily available optimization methods heavily rely on the proper choice of initial values. Furthermore, they do not guarantee convergence to global maxima. Moreover, all model parameters need to be simultaneously maximized, which can be really challenging specifically when dealing with large number of covariates or when having a parameter with respect to which the log-likelihood surface is very flat.$^9$ To overcome these issues with direct maximization of the observed log-likelihood function,$^{27}$ we propose a simplified estimation procedure through the development of a stochastic version of the EM algorithm, called the SEM algorithm. For this purpose, we first need to define the complete data, which includes both observed and missing data. Now, since the number of competing risks $M_i$ related to the occurrence of an event of interest is unknown for each patient in the study, we can treat this as a missing data problem. Thus, we can define the complete data as $\boldsymbol D_c = \{(t_i,\delta_i, M_i, \boldsymbol x_i,\boldsymbol z_i), i = 1,2,\cdots,n\}$. For the cure rate model with the pmf of the number of competing risks as in \eqref{nb}, the joint distribution of $(t_i,\delta_i,m_i)$ is given by
\begin{equation}
f(t_i,\delta_i,m_i;\boldsymbol\theta) = \{S(t_i;\boldsymbol x_i)\}^{m_i - \delta_i} \{m_i f(t_i;\boldsymbol x_i) \}^{\delta_i} p_{m_i}, \ \ t_i>0; \delta_i = 0,1; m_i = \delta_i,\delta_i+1,\cdots.
\label{p1}
\end{equation}
Using  \eqref{p1}, the complete data likelihood function can be expressed as
\begin{eqnarray*}
L_c(\boldsymbol\theta | \boldsymbol D_c) &=& \prod_{i=1}^n f(t_i,\delta_i,m_i;\boldsymbol\theta) = \prod_{i=1}^n \{S(t_i;\boldsymbol x_i)\}^{m_i - \delta_i} \{m_i f(t_i;\boldsymbol x_i) \}^{\delta_i} p_{m_i}.
\end{eqnarray*} 
Hence, the complete data log-likelihood function can be expressed as
\begin{eqnarray}
l_c(\boldsymbol\theta | \boldsymbol D_c) &=& \sum_{i=1}^n \big[ (m_i-\delta_i) \log S(t_i;\boldsymbol x_i) + \delta_i \{  \log m_i + \log f(t_i;\boldsymbol x_i)  \}  + \log p_{m_i} \big] \nonumber \\
&=& \sum_{i=1}^n \bigg[ (m_i-\delta_i) \log S(t_i;\boldsymbol x_i) + \delta_i \log f(t_i;\boldsymbol x_i) + \log \Gamma\bigg(m_i + \frac{1}{\phi}\bigg) - \log \Gamma\bigg(\frac{1}{\phi}\bigg)\nonumber \\
&& + m_i \log \bigg\{ \frac{\phi\eta(\boldsymbol z_i)}{1+\phi\eta(\boldsymbol z_i)} \bigg\} - \frac{1}{\phi} \log (1+\phi\eta(\boldsymbol z_i))  \bigg] +  \sum_{i=1}^n [ \delta_i\log m_i - \log (m_i!) ] \nonumber \\
&=& l_c(\boldsymbol\theta_1) + l_c(\boldsymbol\theta_2) + K,
\label{lc}
\end{eqnarray}
where $\boldsymbol\theta_1=(\phi,\boldsymbol\beta^\prime)^\prime$, $\boldsymbol\theta_2 = (\boldsymbol\alpha^\prime,\boldsymbol\gamma^\prime)^\prime$,
\begin{equation*}
l_c(\boldsymbol\theta_1) =  \sum_{i=1}^n \bigg[ \log \Gamma\bigg(m_i + \frac{1}{\phi}\bigg) - \log \Gamma\bigg(\frac{1}{\phi}\bigg) + m_i \log \bigg\{ \frac{\phi\eta(\boldsymbol z_i)}{1+\phi\eta(\boldsymbol z_i)} \bigg\} - \frac{1}{\phi} \log (1+\phi\eta(\boldsymbol z_i))  \bigg],
\end{equation*}
\begin{equation*}
l_c(\boldsymbol\theta_2) =  \sum_{i=1}^n\big [ (m_i-\delta_i) \log S(t_i;\boldsymbol x_i) + \delta_i \log f(t_i;\boldsymbol x_i) \big],
\end{equation*}
and $K$ is a constant that does not depend on any model parameter, given by
\begin{equation*}
K = \sum_{i=1}^n [ \delta_i\log m_i - \log (m_i!) ].
\end{equation*}
Note that $l_c(\boldsymbol\theta_1)$ involve parameters associated with the cure rate only, whereas $l_c(\boldsymbol\theta_2)$ involve parameters associated with the distribution of the competing risk lifetime only. This facilitates the development of the estimation procedure, which is discussed below. This way of defining the complete log-likelihood function is motivated by a recent work$^{27}$ and is completely different from the standard approach.$^{9,18,20}$

\section{Estimation method: SEM algorithm}

Note that in \eqref{lc}, $m_i's$ are unknown for all $i=1,2,\cdots,n.$ This can be looked as a missing data problem and the well-known EM algorithm can be developed.$^{30}$ However, we develop a stochastic variation of the EM algorithm, called the SEM algorithm.$^{31}$ It is well known that the EM algorithm do not guarantee convergence to a global maximum or even to a local maximum. This is also the case with other Newton-based methods, for instance, the Newton Raphson method. These methods may converge to a stationary point close to the starting value and that stationary point can be a saddle point. Since the SEM algorithm is of stochastic nature, it is free of this saddle point problem.$^{32,33}$ It has been shown that the SEM estimators are efficient under some suitable regulatory conditions.$^{34-36}$ The SEM algorithm is also insensitive to the starting values and it performs well for small and moderate sample sizes.$^{28,33,37}$

In the SEM algorithm, the expectation step (E-step) of the EM algorithm is replaced by a stochastic step (S-step), which is easy to compute as long as the missing data are easy to impute. In the S-step, each missing datum in the complete log-likelihood function is replaced by a value randomly generated from the conditional distribution of the missing data given the observed data and the current values of the parameters. The S-step simulates a pseudo-complete data set and then the maximization step (M-step) involves maximizing the likelihood function based on the complete sample. To facilitate the development of the SEM algorithm, we first need to derive the conditional distribution of $M_i$ given the observed data and parameter values. Now, for the cure rate model with the pmf of the number of competing risks as in \eqref{nb}, the joint distribution of $(t_i,\delta_i)$ is given by
\begin{equation}
f(t_i,\delta_i;\boldsymbol\theta) = \bigg\{\frac{1}{1+\phi\eta(\boldsymbol z_i)F(t_i;\boldsymbol x_i)} \bigg\}^{\frac{1}{\phi}} \bigg\{\frac{f(t_i;\boldsymbol x_i)}{S(t_i;\boldsymbol x_i)} E_{p_i^*}[M_i]   \bigg\}^{\delta_i}, \ \ t_i>0; \delta_i = 0,1,
\label{p2}
\end{equation} 
where $E_{p_i^*}[\cdot]$ denotes that the expectation is taken with respect to a negative binomial distribution with parameters $r=\frac{1}{\phi}$ and $p_i^* = \frac{1+\phi\eta(\boldsymbol z_i)F(t_i;\boldsymbol x_i)}{1+\phi\eta(\boldsymbol z_i)}$. Furthermore, for the negative binomial cure rate model in \eqref{Sp2} and \eqref{fp2}, the conditional distribution of the unobserved number of competing risks $M_i$, i.e., the distribution of $(M_i | t_i,\delta_i;\boldsymbol\theta)$, is given by
\begin{equation}
P[M_i = m_i | t_i,\delta_i; \boldsymbol\theta] =  \frac{m_i^{\delta_i}\frac{\Gamma(m_i + \frac{1}{\phi})}{\Gamma(\frac{1}{\phi})m_i!} \bigg\{ \frac{\phi\eta(\boldsymbol z_i)S(t_i;\boldsymbol x_i)}{1+\phi\eta(\boldsymbol z_i)} \bigg\}^{m_i} \bigg\{ \frac{1+ \phi\eta(\boldsymbol z_i)F(t_i;\boldsymbol x_i)}{1+\phi\eta(\boldsymbol z_i)} \bigg\}^{\frac{1}{\phi}} }{ \bigg\{ \frac{\eta(\boldsymbol z_i) S(t_i;\boldsymbol x_i)}{1+ \phi\eta(\boldsymbol z_i)F(t_i;\boldsymbol x_i) }\bigg\}^{\delta_i}}, \ \ m_i=\delta_i,\delta_i+1,\cdots. 
\label{p3}                   
\end{equation}
From \eqref{p3}, on separating the cases for $\delta_i=0$ and $\delta_i=1$, i.e., for censored and uncensored observations, we have
\begin{equation}
P[M_i=m_i | t_i,\delta_i=0;\boldsymbol\theta] = \frac{\Gamma(m_i + \frac{1}{\phi})}{\Gamma(\frac{1}{\phi})m_i!} \bigg\{ \frac{\phi\eta(\boldsymbol z_i)S(t_i;\boldsymbol x_i)}{1+\phi\eta(\boldsymbol z_i)} \bigg\}^{m_i} \bigg\{ \frac{1+ \phi\eta(\boldsymbol z_i)F(t_i;\boldsymbol x_i)}{1+\phi\eta(\boldsymbol z_i)} \bigg\}^{\frac{1}{\phi}}, \ \ m_i = 0,1,\cdots,
\label{d0}
\end{equation}
which is a negative binomial distribution with parameters $r=\frac{1}{\phi}$ and $p_i^* = \bigg\{ \frac{1+\phi\eta(\boldsymbol z_i)F(t_i;\boldsymbol x_i)}{1+\phi\eta(\boldsymbol z_i)} \bigg\}$. Similarly, we have
\begin{eqnarray}
P[M_i=m_i | t_i,\delta_i=1;\boldsymbol\theta] &=& \frac{m_i\frac{\Gamma(m_i + \frac{1}{\phi})}{\Gamma(\frac{1}{\phi})m_i!} \bigg\{ \frac{\phi\eta(\boldsymbol z_i)S(t_i;\boldsymbol x_i)}{1+\phi\eta(\boldsymbol z_i)} \bigg\}^{m_i} \bigg\{ \frac{1+ \phi\eta(\boldsymbol z_i)F(t_i;\boldsymbol x_i)}{1+\phi\eta(\boldsymbol z_i)} \bigg\}^{\frac{1}{\phi}} }{ \bigg\{ \frac{\eta(\boldsymbol z_i) S(t_i;\boldsymbol x_i)}{1+ \phi\eta(\boldsymbol z_i)F(t_i;\boldsymbol x_i) }\bigg\}}, \ \ m_i = 1,2,\cdots \nonumber \\
&=& \frac{\Gamma(m_i + \frac{1}{\phi})}{\Gamma(\frac{1}{\phi}+1)(m_i-1)!} \bigg\{ \frac{\phi\eta(\boldsymbol z_i)S(t_i;\boldsymbol x_i)}{1+\phi\eta(\boldsymbol z_i)} \bigg\}^{m_i-1} \bigg\{ \frac{1+ \phi\eta(\boldsymbol z_i)F(t_i;\boldsymbol x_i)}{1+\phi\eta(\boldsymbol z_i)} \bigg\}^{\frac{1}{\phi}+1},
\label{d1}
\end{eqnarray}
which is a length-biased negative binomial distribution, length being shifted up by one, with parameters $r=(\frac{1}{\phi}+1)$ and $p_i^* = \bigg\{ \frac{1+\phi\eta(\boldsymbol z_i)F(t_i;\boldsymbol x_i)}{1+\phi\eta(\boldsymbol z_i)} \bigg\}$. 

\subsection{Steps of the SEM algorithm}

{\it Step 1 (Initial guess):} Start with an initial guess of the parameter $\boldsymbol\theta^{(0)}=(\phi^{(0)},\boldsymbol\alpha^{(0)\prime},\boldsymbol\beta^{(0)\prime},\boldsymbol\gamma^{(0)\prime})^\prime$ and the observed data $\boldsymbol D_o$.\\[1ex]
{\it Step 2 (Stochastic step or S-step):} Replace each missing datum $m_i, i=1,2,\cdots,n,$ in the complete data log-likelihood function $l_c(\boldsymbol\theta| \boldsymbol D_c)$ by a value randomly generated from its conditional distribution, given by \eqref{p3}. Thus, for $\delta_i=0$, generate $m_i$ from \eqref{d0}, i.e., generate $m_i$ from a negative binomial distribution with parameters $r=\frac{1}{\phi}$ and $p_i^* = \big\{ \frac{1+\phi\eta(\boldsymbol z_i)F(t_i;\boldsymbol x_i)}{1+\phi\eta(\boldsymbol z_i)} \big\}$, where the parameters are evaluated at $\boldsymbol\theta = \boldsymbol\theta^{(0)}$. Similarly, for $\delta_i=1,$ generate $m_i$ from \eqref{d1}, i.e., first generate $m_i$ from a negative binomial distribution with parameters $r=(\frac{1}{\phi}+1)$ and $p_i^* = \big\{ \frac{1+\phi\eta(\boldsymbol z_i)F(t_i;\boldsymbol x_i)}{1+\phi\eta(\boldsymbol z_i)} \big\}$, where the parameters are evaluated at $\boldsymbol\theta = \boldsymbol\theta^{(0)}$, and, then, replace $m_i$ by $(m_i+1)$. For all $i, i=1,2,\cdots,n,$ denote the generated value of $m_i$ by $\widehat{m_i}^{(0)}$. Replace each unobserved $m_i$ in $\l_c(\boldsymbol\theta | \boldsymbol D_c)$ by $\widehat{m_i}^{(0)}$, and denote the resulting function as
\begin{equation*}
l_c(\boldsymbol\theta;\widehat{\boldsymbol m}^{(0)}) = l_c(\boldsymbol\theta_1;\widehat{\boldsymbol m}^{(0)}) + l_c(\boldsymbol\theta_2;\widehat{\boldsymbol m}^{(0)}) + \widehat{K}^{(0)},
\end{equation*}
where
\begin{equation*}
l_c(\boldsymbol\theta_1;\widehat{\boldsymbol m}^{(0)}) = \sum_{i=1}^n \bigg[ \log \Gamma\bigg(\widehat{m_i}^{(0)} + \frac{1}{\phi}\bigg) - \log \Gamma\bigg(\frac{1}{\phi}\bigg) + \widehat{m_i}^{(0)} \log \bigg\{ \frac{\phi\eta(\boldsymbol z_i)}{1+\phi\eta(\boldsymbol z_i)} \bigg\} - \frac{1}{\phi} \log (1+\phi\eta(\boldsymbol z_i))  \bigg],
\end{equation*}
\begin{equation*}
 l_c(\boldsymbol\theta_2;\widehat{\boldsymbol m}^{(0)}) =  \sum_{i=1}^n\big [ (\widehat{m_i}^{(0)} -\delta_i) \log S(t_i;\boldsymbol x_i) + \delta_i \log f(t_i;\boldsymbol x_i) \big],
\end{equation*}
and
\begin{equation*}
\widehat{K}^{(0)} = \sum_{i=1}^n [ \delta_i\log \widehat{m_i}^{(0)} - \log (\widehat{m_i !}^{(0)} ) ]
\end{equation*}
with $\widehat{\boldsymbol m}^{(0)}$ denoting the vector of $\widehat{m_i}^{(0)}$ values. \\[1ex]
{\it Step 3 (Maximization step or M-step):} Maximize $l_c(\boldsymbol\theta;\widehat{\boldsymbol m}^{(0)})$ with respect to $\boldsymbol\theta$ to find an improved estimate of $\boldsymbol\theta$. This boils down to maximizing 
$l_c(\boldsymbol\theta_1;\widehat{\boldsymbol m}^{(0)})$ with respect to $\boldsymbol\theta_1$ and $l_c(\boldsymbol\theta_2;\widehat{\boldsymbol m}^{(0)})$ with respect to $\boldsymbol\theta_2$, separately. The improved estimates of $\boldsymbol\theta_1$ and $\boldsymbol\theta_2$ are respectively given by
\begin{equation*}
\boldsymbol\theta_1^{(1)} = \operatorname*{arg\, max}_{\boldsymbol\theta_1}l_c(\boldsymbol\theta_1;\widehat{\boldsymbol m}^{(0)}) \ \ \text{and}\ \  \boldsymbol\theta_2^{(1)} = \operatorname*{arg\, max}_{\boldsymbol\theta_2}l_c(\boldsymbol\theta_2;\widehat{\boldsymbol m}^{(0)}).
\end{equation*}
Since the M-step deals with the complete data log-likelihood, it can be easily implemented using standard optimization techniques, for instance, the ``Nelder-Mead'' or ``BFGS'' or ``L-BFGS-B'' methods readily available in R software.\\[1ex]
{\it Step 4 (Iterative step):} Using the updated estimate $\boldsymbol\theta^{(1)} = (\boldsymbol\theta_1^{\prime(1)} , \boldsymbol\theta_2^{\prime(1)} )^\prime$ from Step 3, repeat Step 2 and Step 3 $R$ times to obtain $\boldsymbol\theta^{(k)} , k = 1,2,\cdots,R.$ These sequence of estimates form a Markov chain that do not converge to a single point, but rapidly converges to a stationary distribution, provided some regularity conditions are satisfied.$^{34,38}$ \\[1ex]
{\it Step 5 (Burn-in and MLE):} The aforementioned stationary distribution can be obtained after a burn-in period, and hence the MLE of $\boldsymbol\theta$ can be obtained by discarding the first $r$ iterations for burn-in and averaging over the estimates from the remaining iterations. The random perturbations of the Markov chains prevent the sequence of estimates from being trapped in a local maximum or saddle point, a big advantage of the SEM algorithm.$^{39}$ The MLE of $\boldsymbol\theta$ is finally given by
\begin{eqnarray*}
\hat{\boldsymbol\theta} = \frac{1}{R-r}\sum_{k=r+1}^R\boldsymbol\theta^{(k)}.
\end{eqnarray*}
We denote the above method of finding the MLE as ``MLE (mean)". Another way to calculate the MLE of $\boldsymbol\theta$ is to first calculate the observed data log-likelihood function $l_0(\boldsymbol\theta^{(k)}|D_0)$ for each $k$, $k=r+1,r+2,\cdots,R$, and, then, take the $\boldsymbol\theta^{(k)}$ as the MLE for which the observed data log-likelihood function is the maximum. We denote this method as ``MLE (max log-lik)". A burn-in period of 100 iterations may be sufficient under moderate missing data rates, and an additional 1000 iterations are sufficient to estimate the parameters.$^{40,41}$ However, in our case, a sufficient burn-in period and a sufficient number of iterations will be determined through a preliminary study. Having mentioned this, a trace plot of the sequence of estimates versus the iterations can always be used for validating the sufficiency of the burn-in period, and, if required, a more appropriate burn-in period can be determined. 

In our application, we assume the competing risk lifetime distribution, $W_j, j = 1,2,\cdots,M,$ to follow a Weibull distribution with shape parameter $\frac{1}{\gamma_1}$ and scale parameter $\frac{1}{\gamma_2}$. Thus, the Weibull density function is given by
\begin{equation}
f(t;\boldsymbol x) = \frac{1}{\gamma_1 t} (\gamma_2 t)^{\frac{1}{\gamma_1}}e^{-(\gamma_2 t)^{\frac{1}{\gamma_1}}}, \ \ t > 0, \gamma_1 > 0, \gamma_2 > 0,
\label{ft}
\end{equation}
where $\gamma_2$ is linked to covariate $\boldsymbol x$ through the link function $\gamma_2 = e^{\boldsymbol x^\prime\boldsymbol\alpha}$. Note that one is free to choose any other suitable parametric lifetime distribution or one can also model the competing risk lifetime through the hazard function by using the Cox's proportional model.

\subsection{Advantage over the EM algorithm} 

If we were to develop the EM algorithm under the given framework, the complete data log-likelihood function in \eqref{lc} requires us to calculate the conditional expectations with respect to the terms $\log\bigg\{\Gamma\big(M_i+\frac{1}{\phi}\big)\bigg\}$ and $M_i$.
%\begin{equation}
%E\bigg[ \log\Gamma\big(M_i+\frac{1}{\phi}\big) \big| t_i,\delta_i;\boldsymbol\theta^{(k)}\bigg] \ \ \text{and} \ \ E[M_i | t_i,\delta_i;\boldsymbol\theta^{(k)}],
%\label{EM}
%\end{equation}
%where $\boldsymbol\theta^{(k)}$ denotes the parameter value at the $k-$th iteration step. 
Note that the conditional expectations $E[\log M_i  | t_i,\delta_i;\boldsymbol\theta^{(k)}]$ and $E[\log (M_i!)  | t_i,\delta_i;\boldsymbol\theta^{(k)}]$, coming from the term $K$ in \eqref{lc}, can be ignored since they are independent of any model parameter, and, as such, do not contribute to the maximization step. Now, although calculating the conditional expectation of $M_i$ is not difficult, deriving the conditional expectation with respect to the term $\log\bigg\{\Gamma\big(M_i+\frac{1}{\phi}\big)\bigg\}$ is not straightforward unless the parameter $\phi$ is kept fixed, in which case a profile likelihood approach needs to be employed, in conjunction with the EM algorithm, to estimate the parameter $\phi$.$^{9,27}$ Note that the SEM algorithm easily avoids calculating this complicated conditional expectation as in the case of SEM algorithm it is enough to know the conditional distribution of $M_i$, which in the case of negative binomial competing risks is available in closed form. This can be looked as an advantage of the SEM algorithm over the well-known EM algorithm. This is also the reason why we chose negative binomial competing risks distribution, among several other possible distributions, so as to illustrate the fact that SEM algorithm avoids calculating complicated conditional expectations. In one of the recent works on cure rate model with negative binomial competing risks,$^{42}$ it has been shown how one can avoid the profile likelihood technique by introducing an additional latent variable, say $X$, such that $(M|X=x) \sim Poisson (x)$ and $X\sim Gamma(\frac{1}{\phi},\phi\eta)$. Although this procedure worked well for the real data that was analyzed, the biological interpretation of the continuous latent variable $X$ in the process of disease recurrence or occurrence of an event is not clear. It may be of interest, as a future study, to look at the procedure's finite sample properties and its performance in retrieving the true parameter values for different parameter settings. The proposed SEM algorithm, on the other hand, is based on a simple idea, which is to replace each missing $M_i$ with a value randomly generated from its conditional distribution.  

\subsection{Development of the EM algorithm}

To compare the results of the SEM algorithm with that of the EM algorithm, we also develop the steps of the EM algorithm. For this purpose, we consider the parameter $\phi$ fixed in $l_c(\boldsymbol\theta_1)$ to avoid taking the conditional expectation with respect to the term $\log\big\{\Gamma\big(M_i+\frac{1}{\phi}\big)\big\}$. Thus, $l_c(\boldsymbol\theta_1)$ now reduces to 
\begin{equation*}
l_c(\boldsymbol\theta_1) =  \sum_{i=1}^n \bigg[ m_i \log \bigg\{ \frac{\phi\eta(\boldsymbol z_i)}{1+\phi\eta(\boldsymbol z_i)} \bigg\} - \frac{1}{\phi} \log (1+\phi\eta(\boldsymbol z_i))  \bigg],
\end{equation*}
where $\boldsymbol\theta_1=\boldsymbol\beta$, and the constant $K$ becomes
\begin{equation*}
K = \sum_{i=1}^n \bigg[ \delta_i\log m_i - \log (m_i!) + \log \Gamma\bigg(m_i + \frac{1}{\phi}\bigg) - \log \Gamma\bigg(\frac{1}{\phi}\bigg) \bigg].
\end{equation*}
From \eqref{p3}, it can be shown that at the $k-$th iteration step
\begin{equation*}
E[M_i | t_i,\delta_i;\boldsymbol\theta^{(k)}] = \widehat{m_i}^{(k)} = \left.\frac{\delta_i + \delta_i\phi\eta(\boldsymbol z_i) + \eta(\boldsymbol z_i)S(t_i)}{1+\phi\eta(\boldsymbol z_i)F(t_i)}\right|_{\boldsymbol\theta = \boldsymbol\theta^{(k)}}.
\end{equation*}
Thus, at the $k$-th iteration step, the E-step of the EM algorithm replaces each unobserved $m_i$ in $\l_c(\boldsymbol\theta | \boldsymbol D_c)$ by $\widehat{m_i}^{(k)}$. The resulting function is denoted by
\begin{equation*}
l_c(\boldsymbol\theta;\widehat{\boldsymbol m}^{(k)}) = l_c(\boldsymbol\theta_1;\widehat{\boldsymbol m}^{(k)}) + l_c(\boldsymbol\theta_2;\widehat{\boldsymbol m}^{(k)}) + \widehat{K}^{(k)},
\end{equation*}
where
\begin{equation*}
l_c(\boldsymbol\theta_1;\widehat{\boldsymbol m}^{(k)}) = \sum_{i=1}^n \bigg[ \widehat{m_i}^{(k)} \log \bigg\{ \frac{\phi\eta(\boldsymbol z_i)}{1+\phi\eta(\boldsymbol z_i)} \bigg\} - \frac{1}{\phi} \log (1+\phi\eta(\boldsymbol z_i))  \bigg],
\end{equation*}
\begin{equation*}
 l_c(\boldsymbol\theta_2;\widehat{\boldsymbol m}^{(k)}) =  \sum_{i=1}^n\big [ (\widehat{m_i}^{(k)} -\delta_i) \log S(t_i;\boldsymbol x_i) + \delta_i \log f(t_i;\boldsymbol x_i) \big],
\end{equation*}
and
\begin{equation*}
\widehat{K}^{(k)} = \sum_{i=1}^n \bigg[ \delta_i\log \widehat{m_i}^{(k)} - \log (\widehat{m_i !}^{(k)} ) + \log \Gamma\bigg(\widehat{m_i}^{(k)} + \frac{1}{\phi}\bigg) - \log \Gamma\bigg(\frac{1}{\phi}\bigg)\bigg].
\end{equation*}
In the M-step of the EM algorithm, we maximize the function $l_c(\boldsymbol\theta_1;\widehat{\boldsymbol m}^{(k)})$ and $ l_c(\boldsymbol\theta_2;\widehat{\boldsymbol m}^{(k)})$ with respect to $\boldsymbol\theta_1$ and $\boldsymbol\theta_2$, respectively, to obtain improved estimates of $\boldsymbol\theta_1$ and $\boldsymbol\theta_2$ as
\begin{eqnarray*}
\boldsymbol\theta_1^{(k+1)}= \operatorname*{arg\, max}_{\boldsymbol\theta_1}l_c(\boldsymbol\theta_1;\widehat{\boldsymbol m}^{(k)}).
\end{eqnarray*}
and
\begin{eqnarray*}
\boldsymbol\theta_2^{(k+1)}= \operatorname*{arg\, max}_{\boldsymbol\theta_2}l_c(\boldsymbol\theta_2;\widehat{\boldsymbol m}^{(k)}).
\end{eqnarray*}
We then iterate the E-step and the M-step until we achieve some convergence criterion. For example, we stop the iterative procedure at the $(k+1)$-th iteration step if 
\begin{equation*}
\bigg| \frac{\boldsymbol\theta^{(k+1)} - \boldsymbol\theta^{(k)}}{\boldsymbol\theta^{(k)}}\bigg| < \epsilon,
\end{equation*}
where $\epsilon$ is some desired tolerance. The MLE of $\boldsymbol\theta$ is finally given by $\hat{\boldsymbol\theta} = \boldsymbol\theta^{(k)}$. The parameter $\phi$ (kept constant in the EM algorithm) can be estimated using a profile likelihood technique. In such a technique, we first choose a set of fixed values of $\phi$. Then, for each fixed value of $\phi$, we run the EM algorithm to estimate the model parameters and calculate the value of the observed log-likelihood function. The MLE of $\phi$ is that value of $\phi$ for which the observed log-likelihood function is the maximum. 

\section{Simulation study}

In this section, we demonstrate the performance of the proposed SEM algorithm through the calculated bias, the standard error (SE), the root mean square error (RMSE), and the coverage probabilities (CP) of the asymptotic confidence intervals. For comparison purposes, we also present the corresponding results from the EM algorithm. We consider two different sample sizes, $n=200$ and $n=400$, to study the behavior of the model under varying sample sizes. Then, we divide each sample size into four groups of equal size, where the sample size for each group can be viewed as the number of patients belonging to a particular treatment group or prognostic group. This group category introduces the covariate $x$, where $x=1,2,3,4$. Thus, patients belonging to group $j$ are assigned a covariate value of $x=j, j=1,2,3,4$. We use this covariate to link the parameter $\eta$ associated with the cure rate as well as the parameter $\gamma_2$ associated with the competing risk lifetime distribution. Since for this simulation study, we consider one covariate, we bring in two regression parameter, $\beta_0$ and $\beta_1$, corresponding to the cure rate, and two other regression parameters, $\alpha_0$ and $\alpha_1$, corresponding to the competing risk lifetime distribution. Assuming the cure rate to decrease with an increase in group category, and for a given value of $\phi,$ we can select the cure rates for the first and fourth groups, say $p_{01}$ and $p_{04}$, respectively. Thus, we have
\begin{eqnarray*}
\begin{cases}
p_{01} = \bigg(\frac{1}{1+\phi e^{\beta_0 + \beta_1}}\bigg)^{\frac{1}{\phi}} & \\
p_{04} = \bigg(\frac{1}{1+\phi e^{\beta_0 + 4\beta_1}}\bigg)^{\frac{1}{\phi}} &
\end{cases}
\end{eqnarray*}
\begin{eqnarray} 
\Rightarrow
\begin{cases}
\beta_{1true} = \frac{1}{3}\bigg[ \log\bigg\{\bigg(\frac{1}{p_{04}}\bigg)^\phi-1\bigg\}  - \log\bigg\{\bigg(\frac{1}{p_{01}}\bigg)^\phi-1\bigg\} \bigg] \ \ \text{and} & \\
  \beta_{0true} = \log\bigg\{\bigg(\frac{1}{p_{01}}\bigg)^\phi-1\bigg\} + \log\bigg(\frac{1}{\phi} \bigg)- \beta_{1true}. &
 \end{cases}
 \label{reg}
\end{eqnarray}
We select two different choices, ``High'' and ``Low'', of cure rates for groups 1 and 4 as (0.65,0.25) and (0.40,0.15), respectively. Note that using \eqref{reg}, the cure rates for groups 2 and 3 can be easily calculated as 
\begin{equation*}
p_{02} =  \bigg(\frac{1}{1+\phi e^{ \beta_{0true} + 2  \beta_{1true}}}\bigg)^{\frac{1}{\phi}} \ \ \text{and} \ \  p_{03} =  \bigg(\frac{1}{1+\phi e^{ \beta_{0true} + 3  \beta_{1true}}}\bigg)^{\frac{1}{\phi}}.
\end{equation*}
To incorporate the possibility of censoring corresponding to a susceptible patient, we also fix the overall censoring proportion $p_j$ for the $j-$th group, $j=1,2,3,4$. For this purpose, we select ``High'' and ``Low'' overall censoring proportions as (0.85,0.65,0.50,0.35) and (0.50,0.40,0.30,0.20), respectively. We assume the censoring time $C$ to be random and assume it to follow an exponential distribution with censoring rate $\xi_j$ for group $j, j=1,2,3,4$. Corresponding to the $j-$th group, and for given values of the cure rate $p_{0j}$ and the censoring proportion $p_j$, we can determine $\xi_j$ from the following equation
\begin{eqnarray}
\frac{p_j-p_{0j}}{1-p_{0j}} &=&  \frac{1}{1-p_{0j}} \bigg[ \frac{1}{N} \sum_{i=1}^N \bigg\{ \frac{1}{1+\phi\eta_jF(t_i/\xi_j)} \bigg\}^{\frac{1}{\phi}} - p_{0j} \bigg],
\label{cr}
\end{eqnarray}
which needs to be solved numerically for given values of the model parameters. Note that in \eqref{cr}, $\eta_j=\exp(\beta_0+j \beta_1)$. 

Now, to generate a lifetime for a patient belonging to the $j-$th group, we first generate a censoring time $C$ with the calculated value of censoring rate and a negative binomial competing risk variable $M$ with pmf as in \eqref{nb}. If $M=0,$ it implies that there are no competing risks, and, hence, the lifetime is infinite with respect to the event of interest. As such, in this case, we consider the observed lifetime $T$ to be $C.$ However, if $M>0$, we generate $\{W_1,W_2\cdots,W_M\}$ from the Weibull distribution as given in \eqref{ft}, for chosen values of $\gamma_1$ and $\gamma_2 = e^{\alpha_0+j\alpha_1}$. The observed lifetime is then taken as $T=\min \{Y, C\} \text{, where $Y$ is the minimum of lifetimes } \{W_1,W_2\cdots,W_M\}$. In all cases, if $T=C,$ we set the right censoring indicator variable $\delta=0,$ otherwise, we set $\delta=1.$ Since we assume the cure rate to decrease with an increase in group category, this implies that the expected lifetime will also decrease as we go higher the group category. Thus, the true values of $\gamma_1, \alpha_0,$ and $\alpha_1$ can be selected to achieve a desired mean and variance of the lifetime for a particular group. In our case, we choose $\gamma_1=0.3$, $\alpha_0=-1.5$, and $\alpha_1=0.5$, which ensures that both mean and variance of lifetime decreases with an increase in group category. We also choose two different true values of $\phi$ as 1.5 and 3. In the case of the EM algorithm, to employ the profile likelihood approach to estimate $\phi$, we select the set of $\phi$ as $\{0.1,0.2,\cdots,3 \}$ if the true value of $\phi$ is 1.5, whereas we select the set of $\phi$ as $\{1.5,1.6,\cdots,4.5\}$ if the true value of $\phi$ is 3.

For our simulation study, to find an initial guess of the model parameters, we create an interval for each model parameter by taking 20\% deviation off its true value. Then, for each parameter, we select a value at random from the created interval, which serves as the initial guess. Note that this readily implies that the initial guess for each parameter do not deviate by more than 20\% on either side of the true parameter value, making sure that the initial guess is quite close to the true value. After a preliminary study, we consider the number of SEM iterations $R$ to be 1500 and we use the first 500 iterations as burn-in. For the SEM algorithm, we check both methods of finding the MLEs, that is, MLE (mean) and MLE (max log-lik). However, MLE (mean) results in large bias in the estimate of $\phi$, which is true even for a large sample of size 600. The bias is more noticeable when the cure rates are high. For this reason, for the considered negative binomial cure rate model, we recommend using the method MLE (max log-lik) to find the estimates of the model parameters. All results that we report in our simulation study are averaged over 250 Monte Carlo runs. The R codes for the SEM algorithm are available in the supplementary material of this manuscript.

In Tables \ref{table:T1} and \ref{table:T2}, we present the SEM model fitting results when the cure rates are ``High'' and ``Low'', respectively. We also present the corresponding EM results for the purpose of comparison. First, it is clear that the proposed SEM algorithm performs very well in retrieving the true parameter values for any considered parameter setting. Note that the standard error and RMSE of the estimators decrease with an increase in sample size. The coverage probabilities are also reasonably close to the nominal level used. The results corresponding to the EM algorithm suggest that the profile likelihood approach results in over-coverage of the parameter $\phi$. This is true irrespective of the sample size and the true cure rates. When the true cure rates are high, the over-coverage is noticed for all model parameters except for the parameter $\alpha_0$. In this regard, the performance of the SEM algorithm is much better and hence should be considered as the preferred algorithm. In Table Tables \ref{table:T3} and \ref{table:T4}, we present the estimation results corresponding to the cure rates. Note that irrespective of the true values of the cure rates, these cure rates are estimated with very small bias. The standard error (obtained by using the delta method) and the RMSE of the estimators of cure rates are all small and they further decrease with an increase in sample size. The coverage probabilities are also close to the nominal level. It is interesting to see that the over-coverage the EM algorithm results corresponding to the parameter $\phi$ does not pose any problem to the estimation results for the cure rates, noting that the cure rate is a pure function of $\phi$ and $(\beta_0,\beta_1)$.

In Tables \ref{table:T5} and \ref{table:T6}, we present the SEM results when the method MLE (mean) is used to estimate the model parameters and the cure rates. From Table \ref{table:T5}, it is clear that the bias in the estimate of $\phi$ is very large. Although this bias decreases with an increase in the sample size, even a large sample size of $n=600$ cannot reduce the bias to an extent that can be easily achieved by using the method MLE (max log-lik). Note, however, that this large bias in the estimate of $\phi$ does not pose any problem to the estimation of cure rates, as presented in Table \ref{table:T6}. The results obtained when the true value of $\phi$ is 1.5 are similar and hence are not presented here for the sake of conciseness. 

\textcolor{blue}{To show the superiority of the proposed SEM algorithm, it is also important to compare the performance of the proposed SEM algorithm with two other commonly used estimation procedures; the direct maximization (DM) of the observed log-likelihood function and the Monte Carlo EM (MCEM) algorithm. In Table \ref{table:T7}, we compare the performance of the proposed SEM algorithm with the DM procedure for different parameter settings. For the comparison to be fair, we use the same initial values for both procedures. To employ the DM procedure, we use the ``optim'' function in R with ``BFGS'' method. From the results in Table \ref{table:T7}, note that when the true value of $\phi=3$, the performance of the SEM and DM methods are similar when it comes to the estimation of the parameters associated with the lifetime, i.e., ($\alpha_0,\alpha_1,\gamma_1$). However, when it comes to the estimation of the parameters associated with the cure rate, i.e., $(\beta_0,\beta_1,\phi)$, the SEM, in general, produces smaller bias, SE and RMSE. Note, in particular, the significant reduction in bias and RMSE corresponding to the parameter $\phi$. Thus, the SEM allows more precise estimation of the cure rate. Now, when the true value of $\phi=1.5$, SEM once again results in smaller bias and RMSE. In this case, note that DM results in the coverage probabilities to fall below the nominal level. These findings clearly support the fact that the overall performance of the SEM is better than that of the DM.}

\textcolor{blue}{In Table \ref{table:T8}, we compare the performance of the proposed SEM algorithm with the MCEM algorithm. Once again, we use the same initial values for both SEM and MCEM procedures. To employ the MCEM, we approximate the conditional expectations by the Monte Carlo means based on 500 samples drawn from the conditional distribution of the competing risks. From Table \ref{table:T8}, we can see that in some cases MCEM results in slightly smaller bias, SE and RMSE. However, the MCEM in all cases results in the coverage probabilities to go beyond the nominal level. Since the SEM is based on drawing one sample from the conditional distribution of the missing data along with 1500 iterations, whereas the MCEM is based on drawing multiple samples (taken as 500) in each iteration to approximate the conditional mean, it is also of interest to compare the computing times between these two algorithms. In Table \ref{table:T9}, we present the computing times taken by SEM and MCEM algorithms to produce the estimates of model parameters along with their standard errors for one simulated data. It is clear that the time taken by MCEM is roughly 5 to 6 times the time taken by SEM, implying that MCEM is computationally way more expensive. Given these findings, the proposed SEM algorithm can still be considered as the preferred algorithm.}

\begin{sidewaystable} [!htbp] 
\caption{Comparison of SEM and EM estimation results of model parameters when the true cure rates are high. Note that the SEM results are based on the method MLE (max log-lik).}
\begin{tabular*}{\textwidth}{@{\extracolsep{\fill}} l l l l l l l l l l}\\ 
\hline                                 
$n$ & Parameter & \multicolumn{2}{c}{Estimate (SE)} & \multicolumn{2}{c}{Bias} & \multicolumn{2}{c}{RMSE} & \multicolumn{2}{c}{95\% CP} \\ \cline{3-4}  \cline{5-6} \cline{7-8} \cline{9-10}
& &\multicolumn{1}{c}{SEM}&\multicolumn{1}{c}{EM}&\multicolumn{1}{c}{SEM}& \multicolumn{1}{c}{EM}&\multicolumn{1}{c}{SEM}&\multicolumn{1}{c}{EM}&\multicolumn{1}{c}{SEM}&\multicolumn{1}{c}{EM} \\
\hline  

200 & $\beta_{0}=-1.185$                          &  -$1.308$ ($0.923$) & -$1.279$ ($0.951$) & -$0.123$ & -$0.094$ & $0.874$ & $0.899$ & $0.976$ & $0.980$ \\ 
 &  $\beta_{1}=1.057$                                 & $1.112$  ($0.518$) & $1.148$  ($0.499$) & $0.055$ & $0.090$ & $0.488$ & $0.417$ & $0.956$ & $0.984$ \\ 
 &  $\phi=3$                                                & $2.996$  ($1.859$) & $3.182$  ($1.772$) & -$0.004$ & $0.182$ & $1.574$ & $1.121$ & $0.976$ & $1$ \\ 
 &  $\alpha_0=-1.5$                                    & -$1.501$ ($0.233$) & -$1.500$ ($0.236$) & -$0.001$ & -$0.0003$ & $0.224$ & $0.225$ & $0.948$ & $0.944$ \\ 
 &  $\alpha_1=0.5$                                     & $0.512$  ($0.092$) & $0.504$  ($0.090$) & $0.012$ & $0.004$ & $0.085$ & $0.080$ & $0.956$ & $0.960$ \\ 
 &  $\gamma_1=0.3$                                  & $0.298$  ($0.061$) & $0.290$  ($0.057$) & -$0.002$ & -$0.010$ & $0.052$ & $0.043$ & $0.956$ & $0.996$ \\ [2ex]

  400 & $\beta_{0}=-1.185$ & -$1.212$ ($0.621$) & -$1.211$ ($0.625$) & -$0.028$ & -$0.026$ & $0.635$ & $0.624$ & $0.976$ & $0.972$ \\ 
 &  $\beta_{1}=1.057$ & $1.070$  ($0.334$) & $1.078$  ($0.327$) & $0.013$ & $0.021$ & $0.347$ & $0.310$ & $0.948$ & $0.964$ \\ 
 &  $\phi=3$ &   $2.942$  ($1.211$) & $2.988$  ($1.175$) & -$0.058$ & -$0.012$ & $1.195$ & $0.988$ & $0.972$ & $1$ \\ 
 &  $\alpha_0=-1.5$ &   -$1.493$ ($0.162$) & -$1.493$ ($0.163$) & $0.007$ & $0.007$ & $0.169$ & $0.168$ & $0.932$ & $0.932$ \\ 
 &  $\alpha_1=0.5$  &    $0.506$  ($0.063$) & $0.504$  ($0.063$) & $0.006$ & $0.004$ & $0.064$ & $0.060$ & $0.948$ & $0.960$ \\ 
 &  $\gamma_1=0.3$  & $0.301$  ($0.041$) & $0.299$  ($0.040$) & $0.001$ & -$0.001$ & $0.040$ & $0.035$ & $0.952$ & $0.972$ \\ [2ex]

  200 & $\beta_{0}=-1.182$   & -$1.239$ ($0.697$) & -$1.231$ ($0.699$) & -$0.057$ & -$0.048$ & $0.668$ & $0.666$ & $0.968$ & $0.972$ \\ 
 &  $\beta_{1}=0.681$     &  $0.756$  ($0.399$) & $0.767$  ($0.402$) & $0.076$ & $0.086$ & $0.349$ & $0.310$ & $0.948$ & $0.988$ \\ 
 &  $\phi=1.5$            &   $1.700$  ($1.614$) & $1.766$  ($1.627$) & $0.200$ & $0.266$ & $1.182$ & $0.977$ & $0.952$ & $1$ \\ 
 &  $\alpha_0=-1.5$   &    -$1.488$ ($0.182$) & -$1.491$ ($0.182$) & $0.012$ & $0.009$ & $0.186$ & $0.188$ & $0.936$ & $0.932$ \\ 
 &  $\alpha_1=0.5$   &    $0.498$  ($0.075$) & $0.496$  ($0.075$) & -$0.002$ & -$0.004$ & $0.067$ & $0.064$ & $0.948$ & $0.960$ \\ 
 &  $\gamma_1=0.3$   &   $0.289$  ($0.053$) & $0.286$  ($0.053$) & -$0.011$ & -$0.014$ & $0.044$ & $0.041$ & $0.932$ & $0.964$ \\ [2ex]
 
  400 & $\beta_{0}=-1.182$     &  -$1.196$ ($0.466$) & -$1.194$ ($0.463$) & -$0.013$ & -$0.011$ & $0.477$ & $0.452$ & $0.980$ & $0.984$ \\ 
 &  $\beta_{1}=0.681$             &  $0.728$  ($0.265$) & $0.726$  ($0.264$) & $0.048$ & $0.046$ & $0.255$ & $0.237$ & $0.952$ & $0.976$ \\ 
 &  $\phi=1.5$                          &  $1.652$  ($1.087$) & $1.655$  ($1.090$) & $0.152$ & $0.155$ & $1.024$ & $0.883$ & $0.960$ & $1$ \\ 
 &  $\alpha_0=-1.5$               &   -$1.490$ ($0.124$) & -$1.490$ ($0.124$) & $0.010$ & $0.010$ & $0.126$ & $0.124$ & $0.928$ & $0.932$ \\ 
 &  $\alpha_1=0.5$                  & $0.496$  ($0.052$) & $0.495$  ($0.052$) & -$0.004$ & -$0.005$ & $0.050$ & $0.049$ & $0.952$ & $0.956$ \\ 
 &  $\gamma_1=0.3$             & $0.292$  ($0.037$) & $0.292$  ($0.037$) & -$0.008$ & -$0.008$ & $0.033$ & $0.030$ & $0.948$ & $0.988$ \\ [2ex]

\hline
\end{tabular*}
\label{table:T1}\\
\end{sidewaystable}

\begin{sidewaystable} [!htbp] 
\caption{Comparison of SEM and EM estimation results of model parameters when the true cure rates are low. Note that the SEM results are based on the method MLE (max log-lik)}
\begin{tabular*}{\textwidth}{@{\extracolsep{\fill}} l l l l l l l l l l}\\ 
\hline                                 
$n$ & Parameter & \multicolumn{2}{c}{Estimate (SE)} & \multicolumn{2}{c}{Bias} & \multicolumn{2}{c}{RMSE} & \multicolumn{2}{c}{95\% CP} \\ \cline{3-4}  \cline{5-6} \cline{7-8} \cline{9-10}
& &\multicolumn{1}{c}{SEM}&\multicolumn{1}{c}{EM}&\multicolumn{1}{c}{SEM}& \multicolumn{1}{c}{EM}&\multicolumn{1}{c}{SEM}&\multicolumn{1}{c}{EM}&\multicolumn{1}{c}{SEM}&\multicolumn{1}{c}{EM} \\
\hline  

200 & $\beta_{0}=0.582$ &  $0.515$ ($0.979$)  & $0.637$  ($0.992$) & -$0.067$ & $0.054$  & $0.888$ & $0.912$ & $0.924$ & $0.944$ \\ 
 &  $\beta_{1}=1.002$   &  $1.014$ ($0.411$)  & $1.062$  ($0.400$) & $0.012$  & $0.060$  & $0.386$ & $0.373$ & $0.924$ & $0.968$ \\ 
 &  $\phi=3$                &  $2.943$ ($1.299$)  & $3.180$  ($1.239$) & -$0.057$ & $0.180$  & $1.064$ & $0.965$ & $0.928$ & $0.996$ \\ 
 &  $\alpha_0=-1.5$     &  -$1.456$ ($0.188$) & -$1.467$ ($0.189$) & $0.044$  & $0.033$  & $0.191$ & $0.192$ & $0.952$ & $0.944$ \\ 
 &  $\alpha_1=0.5$    & $0.499$ ($0.075$)  & $0.494$  ($0.076$) & -$0.001$ & -$0.006$ & $0.082$ & $0.080$ & $0.940$ & $0.944$ \\ 
 &  $\gamma_1=0.3$     & $0.302$ ($0.056$)  & $0.291$  ($0.051$) & $0.002$  & -$0.009$ & $0.047$ & $0.042$ & $0.948$ & $0.992$ \\ [2ex]
 
 400 & $\beta_{0}=0.582$     &$0.584$  ($0.673$) & $0.596$  ($0.673$) & $0.002$  & $0.014$ & $0.681$ & $0.666$ & $0.924$ & $0.936$ \\ 
 &  $\beta_{1}=1.002$          &  $1.041$  ($0.270$) & $1.045$  ($0.270$) & $0.039$  & $0.043$ & $0.282$ & $0.259$ & $0.952$ & $0.968$ \\ 
 &  $\phi=3$                          & $3.053$  ($0.839$) & $3.092$  ($0.840$) & $0.053$  & $0.092$ & $0.817$ & $0.773$ & $0.928$ & $0.960$ \\ 
 &  $\alpha_0=-1.5$              & -$1.487$ ($0.131$) & -$1.489$ ($0.131$) & $0.013$  & $0.011$ & $0.138$ & $0.136$ & $0.940$ & $0.948$ \\ 
 &  $\alpha_1=0.5$               & $0.499$  ($0.052$) & $0.498$  ($0.052$) & -$0.001$ & -$0.002$ & $0.059$ & $0.056$ & $0.912$ & $0.944$ \\ 
 &  $\gamma_1=0.3$           &  $0.297$  ($0.036$) & $0.296$  ($0.036$) & -$0.003$ & -$0.004$ & $0.035$ & $0.033$ & $0.936$ & $0.972$ \\ [2ex]
 
  200 & $\beta_{0}=0.110$   &   $0.158$  ($0.687$) & $0.225$  ($0.706$) & $0.048$ & $0.115$ & $0.631$ & $0.664$ & $0.952$ & $0.972$ \\ 
 &  $\beta_{1}=0.568$          &  $0.637$  ($0.301$) & $0.665$  ($0.299$) & $0.070$ & $0.097$ & $0.279$ & $0.282$ & $0.948$ & $0.964$ \\ 
 &  $\phi=1.5$                       & $1.696$  ($1.047$) & $1.846$  ($1.032$) & $0.196$ & $0.346$ & $0.858$ & $0.878$ & $0.976$ & $1$ \\ 
 &  $\alpha_0=-1.5$             &  -$1.487$ ($0.138$) & -$1.493$ ($0.137$) & $0.013$ & $0.007$ & $0.133$ & $0.131$ & $0.940$ & $0.956$ \\ 
 &  $\alpha_1=0.5$              &   $0.494$  ($0.057$) & $0.490$  ($0.057$) & -$0.006$ & -$0.010$ & $0.053$ & $0.053$ & $0.960$ & $0.968$ \\ 
 &  $\gamma_1=0.3$            &  $0.288$  ($0.047$) & $0.282$  ($0.045$) & -$0.012$ & -$0.018$ & $0.042$ & $0.043$ & $0.940$ & $0.968$ \\ [2ex]
 
 400 & $\beta_{0}=0.110$     &  $0.130$  ($0.449$) & $0.157$  ($0.458$) & $0.020$ & $0.047$ & $0.455$ & $0.461$ & $0.936$ & $0.944$ \\ 
 &  $\beta_{1}=0.568$           & $0.580$  ($0.196$) & $0.587$  ($0.196$) & $0.012$ & $0.019$ & $0.184$ & $0.189$ & $0.956$ & $0.948$ \\ 
 &  $\phi=1.5$                      & $1.511$  ($0.692$) & $1.570$  ($0.699$) & $0.011$ & $0.070$ & $0.618$ & $0.637$ & $0.956$ & $0.980$ \\ 
 &  $\alpha_0=-1.5$             & -$1.492$ ($0.096$) & -$1.495$ ($0.097$) & $0.008$ & $0.005$ & $0.100$ & $0.099$ & $0.956$ & $0.964$ \\ 
 &  $\alpha_1=0.5$               &  $0.499$  ($0.040$) & $0.498$  ($0.040$) & -$0.001$ & -$0.002$ & $0.038$ & $0.038$ & $0.940$ & $0.936$ \\ 
 &  $\gamma_1=0.3$             &  $0.298$  ($0.033$) & $0.296$  ($0.032$) & -$0.002$ & -$0.004$ & $0.030$ & $0.030$ & $0.976$ & $0.980$ \\ [2ex]
\hline
\end{tabular*}
\label{table:T2}\\
\end{sidewaystable}

\begin{sidewaystable}  [!htbp] 
\caption{Comparison of SEM and EM estimation results for cure rates when the true cure rates are high. Note that the SEM results are based on the method MLE (max log-lik)}
\begin{tabular*}{\textwidth}{@{\extracolsep{\fill}} l l l l l l l l l l}\\ 
\hline                                 
$n$ & Parameter & \multicolumn{2}{c}{Estimate (SE)} & \multicolumn{2}{c}{Bias} & \multicolumn{2}{c}{RMSE} & \multicolumn{2}{c}{95\% CP} \\ \cline{3-4}  \cline{5-6} \cline{7-8} \cline{9-10}
& &\multicolumn{1}{c}{SEM}&\multicolumn{1}{c}{EM}&\multicolumn{1}{c}{SEM}& \multicolumn{1}{c}{EM}&\multicolumn{1}{c}{SEM}&\multicolumn{1}{c}{EM}&\multicolumn{1}{c}{SEM}&\multicolumn{1}{c}{EM} \\
\hline  
200 & $p_{01}=0.650$      & $0.660$ ($0.088$) & $0.659$ ($0.088$) & $0.010$ & $0.009$ & $0.087$ & $0.088$ & $0.928$ & $0.928$ \\ 
 &  $p_{02}=0.488$            &$0.496$ ($0.056$) & $0.494$ ($0.055$) & $0.008$ & $0.006$ & $0.059$ & $0.058$ & $0.940$ & $0.932$ \\ 
 &  $p_{03}=0.352$          & $0.354$ ($0.046$) & $0.353$ ($0.046$) & $0.002$ & $0.001$ & $0.051$ & $0.048$ & $0.936$ & $0.948$ \\ 
 &  $p_{04}=0.250$          &  $0.247$ ($0.055$) & $0.249$ ($0.055$) & -$0.003$ & -$0.001$ & $0.057$ & $0.053$ & $0.928$ & $0.952$ \\ [2ex]
 
400 & $p_{01}=0.650$     &  $0.651$  ($0.062$) & $0.651$ ($0.063$) & $0.001$ & $0.001$ & $0.064$ & $0.064$ & $0.940$ & $0.948$ \\ 
 &  $p_{02}=0.488$          & $0.489$  ($0.039$) & $0.488$ ($0.039$) & $0.001$ & $0.0004$ & $0.039$ & $0.038$ & $0.948$ & $0.952$ \\ 
 &  $p_{03}=0.352$           & $0.350$  ($0.032$) & $0.350$ ($0.032$) & -$0.003$ & -$0.003$ & $0.031$ & $0.030$ & $0.960$ & $0.972$ \\ 
 &  $p_{04}=0.250$         &  $0.245$  ($0.039$) & $0.246$ ($0.039$) & -$0.005$ & -$0.004$ & $0.037$ & $0.036$ & $0.972$ & $0.964$ \\ [2ex]
 
 200 & $p_{01}=0.650$      &  $0.652$ ($0.084$) & $0.652$ ($0.084$) & $0.002$ & $0.002$ & $0.086$ & $0.085$ & $0.928$ & $0.924$ \\ 
 &  $p_{02}=0.504$            & $0.507$ ($0.057$) & $0.506$ ($0.056$) & $0.003$ & $0.002$ & $0.060$ & $0.059$ & $0.924$ & $0.940$ \\ 
&  $p_{03}=0.364$             & $0.365$ ($0.046$) & $0.364$ ($0.045$) & $0.001$ & -$0.0004$ & $0.044$ & $0.043$ & $0.956$ & $0.956$ \\ 
 &  $p_{04}=0.250$            &  $0.249$ ($0.058$) & $0.249$ ($0.057$) & -$0.001$ & -$0.001$ & $0.052$ & $0.050$ & $0.976$ & $0.980$ \\ [2ex]
 
 400 & $p_{01}=0.650$      & $0.650$ ($0.059$) & $0.649$ ($0.059$) & -$0.0001$ & -$0.001$ & $0.059$ & $0.058$ & $0.948$ & $0.948$ \\ 
 &  $p_{02}=0.504$            & $0.504$ ($0.040$) & $0.503$ ($0.039$) & -$0.0003$ & -$0.001$ & $0.039$ & $0.039$ & $0.936$ & $0.928$ \\ 
 &  $p_{03}=0.364$             & $0.363$ ($0.032$) & $0.364$ ($0.032$) & -$0.001$ & -$0.001$ & $0.036$ & $0.035$ & $0.908$ & $0.916$ \\ 
 &  $p_{04}=0.250$            & $0.248$ ($0.040$) & $0.249$ ($0.041$) & -$0.002$ & -$0.001$ & $0.045$ & $0.045$ & $0.908$ & $0.932$ \\ [2ex]
\hline
\end{tabular*}
\label{table:T3}\\
\end{sidewaystable}

\begin{sidewaystable}  [!htbp] 
\caption{Comparison of SEM and EM estimation results for cure rates when the true cure rates are low. Note that the SEM results are based on the method MLE (max log-lik)}
\begin{tabular*}{\textwidth}{@{\extracolsep{\fill}} l l l l l l l l l l}\\ 
\hline                                 
$n$ & Parameter & \multicolumn{2}{c}{Estimate (SE)} & \multicolumn{2}{c}{Bias} & \multicolumn{2}{c}{RMSE} & \multicolumn{2}{c}{95\% CP} \\ \cline{3-4}  \cline{5-6} \cline{7-8} \cline{9-10}
& &\multicolumn{1}{c}{SEM}&\multicolumn{1}{c}{EM}&\multicolumn{1}{c}{SEM}& \multicolumn{1}{c}{EM}&\multicolumn{1}{c}{SEM}&\multicolumn{1}{c}{EM}&\multicolumn{1}{c}{SEM}&\multicolumn{1}{c}{EM} \\
\hline  
200 & $p_{01}=0.400$   &$0.410$ ($0.063$) & $0.408$ ($0.062$) & $0.010$  & $0.008$ & $0.064$ & $0.063$ & $0.948$ & $0.924$ \\ 
 &  $p_{02}=0.290$        &  $0.293$ ($0.037$) & $0.293$ ($0.037$) & $0.003$  & $0.003$ & $0.035$ & $0.034$ & $0.948$ & $0.952$ \\ 
 &  $p_{03}=0.209$        &  $0.208$ ($0.035$) & $0.210$ ($0.035$) & -$0.001$ & $0.001$ & $0.033$ & $0.032$ & $0.948$ & $0.952$ \\ 
 &  $p_{04}=0.150$        &  $0.149$ ($0.038$) & $0.151$ ($0.038$) & -$0.001$ & $0.001$ & $0.038$ & $0.036$ & $0.928$ & $0.944$ \\ [2ex]
 
400 & $p_{01}=0.400$    &  $0.404$ ($0.044$) & $0.404$ ($0.044$) & $0.004$ & $0.004$ & $0.047$ & $0.046$ & $0.932$ & $0.936$ \\ 
 &  $p_{02}=0.290$         & $0.290$ ($0.026$) & $0.291$ ($0.026$) & -$0.001$ & $0.0001$ & $0.026$ & $0.025$ & $0.944$ & $0.940$ \\ 
 &  $p_{03}=0.209$          & $0.206$ ($0.025$) & $0.207$ ($0.025$) & -$0.003$ & -$0.002$ & $0.023$ & $0.024$ & $0.960$ & $0.960$ \\ 
 &  $p_{04}=0.150$          &  $0.147$ ($0.027$) & $0.148$ ($0.027$) & -$0.003$ & -$0.002$ & $0.025$ & $0.026$ & $0.940$ & $0.940$ \\ [2ex]
 
200 & $p_{01}=0.400$      &   $0.405$ ($0.064$) & $0.404$ ($0.064$) & $0.005$ & $0.004$ & $0.066$ & $0.065$ & $0.944$ & $0.940$ \\ 
 &  $p_{02}=0.296$           & $0.297$ ($0.038$) & $0.297$ ($0.038$) & $0.001$ & $0.001$ & $0.040$ & $0.039$ & $0.944$ & $0.944$ \\ 
 &  $p_{03}=0.213$            &  $0.212$ ($0.035$) & $0.214$ ($0.035$) & -$0.001$ & $0.001$ & $0.035$ & $0.034$ & $0.960$ & $0.968$ \\ 
 &  $p_{04}=0.150$            & $0.150$ ($0.041$) & $0.152$ ($0.041$) & -$0.0003$ & $0.002$ & $0.037$ & $0.037$ & $0.960$ & $0.964$ \\ [2ex]
 
400 & $p_{01}=0.400$      & $0.399$ ($0.046$) & $0.399$ ($0.045$) & -$0.001$ & -$0.001$ & $0.048$ & $0.047$ & $0.936$ & $0.944$ \\ 
 &  $p_{02}=0.296$            &  $0.294$ ($0.027$) & $0.295$ ($0.027$) & -$0.002$ & -$0.001$ & $0.027$ & $0.027$ & $0.952$ & $0.956$ \\ 
 &  $p_{03}=0.213$           &$0.210$ ($0.025$) & $0.212$ ($0.025$) & -$0.002$ & -$0.001$ & $0.025$ & $0.025$ & $0.948$ & $0.940$ \\ 
 &  $p_{04}=0.150$            &$0.148$ ($0.029$) & $0.150$ ($0.029$) & -$0.002$ & -$0.0002$ & $0.029$ & $0.029$ & $0.956$ & $0.952$ \\ [2ex]
\hline
\end{tabular*}
\label{table:T4}\\
\end{sidewaystable}

\begin{table}  [!htbp] 
\caption{SEM estimation results for model parameters when the method MLE (mean) is used. }
\begin{tabular*}{\textwidth}{@{\extracolsep{\fill}} l l l l l l l}\\ 
\hline                                 
$n$ & Cure Rate & Parameter & Estimate (SE) & Bias & RMSE & 95\% CP \\  
\hline   
  200 & High & $\beta_{0}=-1.185$                         & -1.236 (0.731) & -0.052 & 0.630 & 0.988 \\
 &  & $\beta_{1}=1.057$                               & 0.728 (0.547) & -0.329 & 0.452 & 0.900  \\
 &  & $\phi=3$                                    & 1.470 (2.215) & -1.530 & 1.844 & 0.924  \\
 &  & $\alpha_0=-1.5$                        & -1.500 (0.226) & 0.000 & 0.228 & 0.936  \\
 &  & $\alpha_1=0.5$                         & 0.559 (0.107) & 0.059 & 0.103 & 0.932 \\ 
 &  & $\gamma_1=0.3$                      & 0.347 (0.095) & 0.047 & 0.066 & 0.988 \\ [2ex]
 
  400 & High & $\beta_{0}=-1.185$                           & -1.216 (0.533) & -0.031 & 0.500 & 0.992 \\
 & & $\beta_{1}=1.057$                               & 0.872 (0.364) & -0.185 & 0.320 & 0.920  \\
 &  & $\phi=3$                                    & 2.138 (1.396) & -0.861 & 1.285 & 0.924  \\
 &  & $\alpha_0=-1.5$                        & -1.498 (0.160) & 0.001 & 0.153 & 0.940  \\
 &  & $\alpha_1=0.5$                         & 0.529 (0.070) & 0.029 & 0.066 & 0.944 \\ 
 &  & $\gamma_1=0.3$                      & 0.331 (0.057) & 0.031 & 0.048 & 0.984 \\ [2ex]
 
 600 & High & $\beta_{0}=-1.185$                           & -1.187 (0.458) & -0.003  & 0.447 & 0.976 \\
 & & $\beta_{1}=1.057$                               & 0.931 (0.271) & -0.126  & 0.280 & 0.908  \\
 &  & $\phi=3$                                    & 2.473 (1.025) & -0.526 & 1.045 & 0.900  \\
 &  & $\alpha_0=-1.5$                        & -1.515 (0.131) & -0.015 & 0.132 & 0.940  \\
 &  & $\alpha_1=0.5$                         & 0.524 (0.053) & 0.024 & 0.058 & 0.932 \\ 
 &  & $\gamma_1=0.3$                      & 0.319 (0.039) & 0.019 & 0.037 & 0.964 \\ [2ex]

  200 & Low & $\beta_{0}=0.582$                         & 0.025 (0.804) & -0.558 & 0.822 & 0.796 \\
 &  & $\beta_{1}=1.002$                               & 0.750 (0.435) & -0.251 & 0.394 & 0.892  \\
 &  & $\phi=3$                                    & 1.800 (1.509) & -1.200 & 1.453 & 0.864  \\
 &  & $\alpha_0=-1.5$                        & -1.430 (0.187) & 0.070 & 0.176 & 0.956  \\
 &  & $\alpha_1=0.5$                         & 0.536 (0.084) & 0.036 & 0.084 & 0.920 \\ 
 &  & $\gamma_1=0.3$                      & 0.358 (0.092) & 0.058 & 0.075 & 0.972 \\ [2ex]
 
  400 & Low & $\beta_{0}=0.582$                           & 0.312 (0.613) & -0.271 & 0.569 & 0.908 \\
 &  & $\beta_{1}=1.002$                               & 0.846 (0.272) & -0.156 & 0.275 & 0.920  \\
 &  & $\phi=3$                                    & 2.332 (0.920) & -0.668 & 0.927 & 0.904  \\
 &  & $\alpha_0=-1.5$                        & -1.465 (0.132) & 0.035 & 0.120 & 0.960  \\
 &  & $\alpha_1=0.5$                         & 0.519 (0.053) & 0.019 & 0.056 & 0.928 \\ 
 &  & $\gamma_1=0.3$                      & 0.331 (0.049) & 0.031 & 0.045 & 0.980 \\ [2ex]
 
 600 & Low & $\beta_{0}=0.582$                           & 0.414 (0.501) & -0.168  & 0.486 & 0.904 \\
 & & $\beta_{1}=1.002$                               & 0.873 (0.214) & -0.128  & 0.223 & 0.916  \\
 &  & $\phi=3$                                    & 2.525 (0.706) & -0.475 & 0.741 & 0.884  \\
 &  & $\alpha_0=-1.5$                        & -1.481 (0.107) & 0.019 & 0.107 & 0.952  \\
 &  & $\alpha_1=0.5$                         & 0.514 (0.043) & 0.014 & 0.043 & 0.940 \\ 
 &  & $\gamma_1=0.3$                      & 0.324 (0.036) & 0.024 & 0.037 & 0.964 \\ [2ex]

\hline
\end{tabular*}
\label{table:T5}\\
\end{table}

\begin{table}  [!htbp] 
\caption{SEM estimation results for cure rates when the method MLE (mean) is used. }
\begin{tabular*}{\textwidth}{@{\extracolsep{\fill}} l l l l l l l}\\ 
\hline                                 
$n$  & Cure Rate & Parameter & Estimate (SE) & Bias & RMSE & 95\% CP \\  
\hline   
 200 & High & $p_{01}=0.650$                          & 0.646 (0.090) & -0.004  & 0.087 & 0.940 \\
 &  & $p_{02}=0.488$                               & 0.498 (0.060) & 0.010  & 0.060 & 0.952  \\
 &  & $p_{03}=0.352$                               & 0.352 (0.048) & 0.000 & 0.042 & 0.960 \\
 & &  $p_{04}=0.250$                               & 0.234 (0.156) & -0.016  & 0.053 & 0.940  \\ [2ex]
 
 400 & High & $p_{01}=0.650$                           & 0.649 (0.062) & -0.001  & 0.061 & 0.956 \\
 & & $p_{02}=0.488$                               & 0.492 (0.041) & 0.004  & 0.042 & 0.936  \\
 & & $p_{03}=0.352$                               & 0.350 (0.032) & -0.002 & 0.034 & 0.928 \\
 & & $p_{04}=0.250$                               & 0.239 (0.039) & -0.011  & 0.040 & 0.916  \\ [2ex]
 
 600 & High & $p_{01}=0.650$                           & 0.649 (0.051) & -0.001  & 0.051 & 0.964 \\
 &  & $p_{02}=0.488$                               & 0.492 (0.032) & 0.004  & 0.033 & 0.944  \\
 & &  $p_{03}=0.352$                               & 0.353 (0.026) & 0.001 & 0.028 & 0.920 \\
 & &  $p_{04}=0.250$                               & 0.247 (0.032) & -0.003  & 0.034 & 0.940  \\ [2ex]
  
200 & Low & $p_{01}=0.400$                           & 0.418 (0.066) & 0.018  & 0.066 & 0.928 \\
 & &  $p_{02}=0.290$                               & 0.295 (0.039) & 0.005  & 0.038 & 0.944  \\
 & & $p_{03}=0.209$                               & 0.202 (0.034) & -0.007 & 0.036 & 0.912 \\
 & & $p_{04}=0.150$                               & 0.137 (0.038) & -0.013  & 0.042& 0.856  \\ [2ex]
 
 400 & Low & $p_{01}=0.400$                           & 0.405 (0.046) & 0.005  & 0.045 & 0.944 \\
 & & $p_{02}=0.290$                               & 0.290 (0.026) & 0.000  & 0.027 & 0.956  \\
 & & $p_{03}=0.209$                               & 0.205 (0.024) & -0.004 & 0.025 & 0.932 \\
 & & $p_{04}=0.150$                               & 0.144 (0.027) & -0.006  & 0.027 & 0.920  \\ [2ex]
 
 600 & Low & $p_{01}=0.400$                           & 0.402 (0.037) & 0.002  & 0.036 & 0.968 \\
 & &  $p_{02}=0.290$                               & 0.291 (0.021) & 0.000  & 0.021 & 0.952  \\
 & & $p_{03}=0.209$                               & 0.208 (0.020) & -0.001 & 0.020 & 0.932 \\
 & & $p_{04}=0.150$                               & 0.148 (0.022) & -0.002  & 0.022 & 0.924  \\ [2ex]

\hline
\end{tabular*}
\label{table:T6}\\
\end{table}

\begin{sidewaystable} [!htbp] 
\color{blue}
\fontsize{10}{10}\selectfont
\caption{\textcolor{blue}{Comparison of SEM and DM estimation results of model parameters under different parameter settings}}
\begin{tabular*}{\textwidth}{@{\extracolsep{\fill}} l l l l l l l l l l l}\\ 
\hline                                 
$n$ & Cure Rate & Parameter & \multicolumn{2}{c}{Estimate (SE)} & \multicolumn{2}{c}{Bias} & \multicolumn{2}{c}{RMSE} & \multicolumn{2}{c}{95\% CP} \\ \cline{4-5}  \cline{6-7} \cline{8-9} \cline{10-11}
& & &\multicolumn{1}{c}{SEM}&\multicolumn{1}{c}{DM}&\multicolumn{1}{c}{SEM}& \multicolumn{1}{c}{DM}&\multicolumn{1}{c}{SEM}&\multicolumn{1}{c}{DM}&\multicolumn{1}{c}{SEM}&\multicolumn{1}{c}{DM} \\
\hline  

200 & High & $\beta_{0}=-1.185$ &         $$-$1.226$ ($0.935$) & $$-$1.153$ ($1.006$) & $$-$0.041$ & $0.032$ & $0.876$ & $1.002$ & $0.976$ & $0.956$ \\ 
 &  & $\beta_{1}=1.057	$   &            $1.086$ ($0.507$) & $1.188$ ($0.510$) & $0.028$ & $0.130$ & $0.440$ & $0.547$ & $0.936$ & $0.952$ \\ 
 &  & $\phi=3$                &              $3.021$ ($1.877$) & $3.493$ ($1.880$) & $0.021$ & $0.493$ & $1.507$ & $2.083$ & $0.956$ & $0.968$ \\  
 &  & $\alpha_0=-1.5$     &            $$-$1.497$ ($0.234$) & $$-$1.498$ ($0.232$) & $0.003$ & $0.002$ & $0.239$ & $0.245$ & $0.956$ & $0.924$ \\ 
 &  & $\alpha_1=0.5$    &              $0.511$ ($0.091$) & $0.503$ ($0.089$) & $0.011$ & $0.003$ & $0.091$ & $0.094$ & $0.948$ & $0.932$ \\ 
 & &  $\gamma_1=0.3$     &          $0.297$ ($0.062$) & $0.286$ ($0.055$) & $$-$0.003$ & $$-$0.014$ & $0.050$ & $0.059$ & $0.944$ & $0.900$ \\ [2ex]
 
 400 & High & $\beta_{0}=-1.185$     &           $$-$1.207$ ($0.632$) & $$-$1.179$ ($0.653$) & $$-$0.023$ & $0.005$ & $0.652$ & $0.679$ & $0.956$ & $0.960$ \\ 
 &  & $\beta_{1}=1.057$          &            $1.094$ ($0.337$) & $1.120$ ($0.336$) & $0.037$ & $0.063$ & $0.327$ & $0.338$ & $0.948$ & $0.956$ \\
 &  & $\phi=3$                          &            $3.090$ ($1.220$) & $3.232$ ($1.227$) & $0.090$ & $0.232$ & $1.096$ & $1.228$ & $0.976$ & $0.972$ \\
 &  & $\alpha_0=-1.5$              &           $$-$1.490$ ($0.161$) & $$-$1.493$ ($0.162$) & $0.010$ & $0.007$ & $0.186$ & $0.183$ & $0.900$ & $0.916$ \\ 
 &  & $\alpha_1=0.5$               &            $0.502$ ($0.062$) & $0.500$ ($0.062$) & $0.002$ & $$-$0.0003$ & $0.070$ & $0.069$ & $0.924$ & $0.920$ \\
 &  & $\gamma_1=0.3$           &            $0.295$ ($0.041$) & $0.292$ ($0.040$) & $$-$0.005$ & $$-$0.008$ & $0.034$ & $0.036$ & $0.968$ & $0.948$ \\ [2ex]
 
%  200 & $\beta_{0}=0.582$   &                $0.537$ ($1.149$) & $0.726$ ($1.012$) & $$-$0.045$ & $0.144$ & $0.854$ & $1.043$ & $0.936$ & $0.932$ \\
 %&  $\beta_{1}=1.002$          &                $0.972$ ($0.407$) & $1.050$ ($0.396$) & $$-$0.030$ & $0.048$ & $0.363$ & $0.405$ & $0.924$ & $0.940$ \\
 %&  $\phi=3$                       &                $2.831$ ($1.465$) & $3.176$ ($1.251$) & $$-$0.169$ & $0.176$ & $1.062$ & $1.353$ & $0.948$ & $0.952$ \\ 
 %&  $\alpha_0=-1.5$             &                $$-$1.480$ ($0.205$) & $$-$1.493$ ($0.187$) & $0.020$ & $0.007$ & $0.183$ & $0.183$ & $0.952$ & $0.956$ \\
 %&  $\alpha_1=0.5$              &                $0.511$ ($0.075$) & $0.506$ ($0.075$) & $0.011$ & $0.006$ & $0.080$ & $0.079$ & $0.936$ & $0.944$ \\ 
 %&  $\gamma_1=0.3$            &               $0.304$ ($0.065$) & $0.291$ ($0.050$) & $0.004$ & $$-$0.009$ & $0.047$ & $0.052$ & $0.980$ & $0.928$ \\ [2ex]
 
 400 & Low & $\beta_{0}=0.582$     &             $0.633$ ($0.675$) & $0.706$ ($0.690$) & $0.051$ & $0.124$ & $0.710$ & $0.798$ & $0.948$ & $0.940$ \\
 & & $\beta_{1}=1.002$           &             $1.004$ ($0.265$) & $1.035$ ($0.270$) & $0.002$ & $0.034$ & $0.263$ & $0.285$ & $0.948$ & $0.948$ \\ 
 &  & $\phi=3$                      &               $3.040$ ($0.841$) & $3.174$ ($0.850$) & $0.040$ & $0.174$ & $0.774$ & $0.933$ & $0.964$ & $0.944$ \\ 
 &  & $\alpha_0=-1.5$             &               $$-$1.501$ ($0.131$) & $$-$1.505$ ($0.131$) & $$-$0.001$ & $$-$0.005$ & $0.130$ & $0.131$ & $0.948$ & $0.940$ \\ 
 &  & $\alpha_1=0.5$               &              $0.506$ ($0.052$) & $0.503$ ($0.052$) & $0.006$ & $0.003$ & $0.051$ & $0.051$ & $0.952$ & $0.944$ \\
 &  & $\gamma_1=0.3$             &           $0.298$ ($0.036$) & $0.294$ ($0.035$) & $$-$0.002$ & $$-$0.006$ & $0.036$ & $0.039$ & $0.940$ & $0.908$ \\ [2ex]
 
  200 & High & $\beta_{0}=-1.182$     &             $$-$1.251$ ($0.722$) & $$-$1.222$ ($0.753$) & $$-$0.069$ & $$-$0.040$ & $0.712$ & $0.814$ & $0.988$ & $0.908$ \\ 
 &  & $\beta_{1}=0.681$           &             $0.786$ ($0.407$) & $0.827$ ($0.393$) & $0.105$ & $0.147$ & $0.397$ & $0.462$ & $0.956$ & $0.896$ \\ 
 &  & $\phi=1.5$                      &               $1.829$ ($1.642$) & $2.009$ ($1.599$) & $0.329$ & $0.509$ & $1.398$ & $1.928$ & $0.940$ & $0.904$ \\
 &  & $\alpha_0=-1.5$             &               $$-$1.498$ ($0.182$) & $$-$1.496$ ($0.171$) & $0.002$ & $0.004$ & $0.175$ & $0.176$ & $0.952$ & $0.860$ \\
 &  & $\alpha_1=0.5$               &              $0.498$ ($0.075$) & $0.496$ ($0.068$) & $$-$0.002$ & $$-$0.004$ & $0.068$ & $0.076$ & $0.932$ & $0.868$ \\
 &  & $\gamma_1=0.3$             &           $0.284$ ($0.053$) & $0.281$ ($0.047$) & $$-$0.016$ & $$-$0.019$ & $0.046$ & $0.055$ & $0.928$ & $0.832$ \\ [2ex]
 
   400 & High & $\beta_{0}=-1.182$     &          $$-$1.174$ ($0.460$) & $$-$1.161$ ($0.456$) & $0.009$ & $0.022$ & $0.472$ & $0.489$ & $0.976$ & $0.928$ \\ 
 & & $\beta_{1}=0.681$           &             $0.690$ ($0.259$) & $0.706$ ($0.251$) & $0.010$ & $0.025$ & $0.255$ & $0.266$ & $0.936$ & $0.900$ \\ 
 & & $\phi=1.5$                      &               $1.525$ ($1.112$) & $1.606$ ($1.064$) & $0.025$ & $0.106$ & $0.976$ & $1.086$ & $0.972$ & $0.940$ \\ 
 & & $\alpha_0=-1.5$             &              $$-$1.503$ ($0.125$) & $$-$1.504$ ($0.119$) & $$-$0.003$ & $$-$0.004$ & $0.140$ & $0.139$ & $0.924$ & $0.880$ \\
 & & $\alpha_1=0.5$               &             $0.504$ ($0.052$) & $0.502$ ($0.049$) & $0.004$ & $0.002$ & $0.051$ & $0.052$ & $0.924$ & $0.896$ \\ 
 & & $\gamma_1=0.3$             &           $0.295$ ($0.037$) & $0.294$ ($0.035$) & $$-$0.005$ & $$-$0.006$ & $0.032$ & $0.035$ & $0.976$ & $0.916$ \\ [2ex]
 
\hline
\end{tabular*}
\label{table:T7}\\
\end{sidewaystable}

\begin{sidewaystable} [!htbp] 
\color{blue}
\fontsize{10}{10}\selectfont
\caption{\textcolor{blue}{Comparison of SEM and MCEM estimation results of model parameters under different parameter settings}}
\begin{tabular*}{\textwidth}{@{\extracolsep{\fill}} l l l l l l l l l l l}\\ 
\hline                                 
$n$ & Cure Rate & Parameter & \multicolumn{2}{c}{Estimate (SE)} & \multicolumn{2}{c}{Bias} & \multicolumn{2}{c}{RMSE} & \multicolumn{2}{c}{95\% CP} \\ \cline{4-5}  \cline{6-7} \cline{8-9} \cline{10-11}
& & &\multicolumn{1}{c}{SEM}&\multicolumn{1}{c}{MCEM}&\multicolumn{1}{c}{SEM}& \multicolumn{1}{c}{MCEM}&\multicolumn{1}{c}{SEM}&\multicolumn{1}{c}{MCEM}&\multicolumn{1}{c}{SEM}&\multicolumn{1}{c}{MCEM} \\
\hline  

200 & High & $\beta_{0}=-1.185$ &               $$-$1.166$ ($1.027$) & $$-$1.161$ ($0.940$) & $0.018$ & $0.024$ & $0.900$ & $0.914$ & $0.984$ & $0.992$ \\ 	
 &  & $\beta_{1}=1.057$   &                          $1.021$ ($0.550$) & $1.058$ ($0.484$) & $$-$0.036$ & $0.001$ & $0.448$ & $0.345$ & $0.924$ & $0.972$ \\
 &  & $\phi=3$                &                              $2.885$ ($2.192$) & $3.091$ ($1.783$) & $$-$0.115$ & $0.091$ & $1.478$ & $0.957$ & $0.940$ & $0.996$ \\
 &  & $\alpha_0=-1.5$     &                            $$-$1.484$ ($0.238$) & $$-$1.481$ ($0.234$) & $0.016$ & $0.019$ & $0.239$ & $0.236$ & $0.944$ & $0.952$ \\
 &  & $\alpha_1=0.5$    &                                $0.513$ ($0.090$) & $0.504$ ($0.089$) & $0.013$ & $0.004$ & $0.090$ & $0.081$ & $0.948$ & $0.956$ \\
 & &  $\gamma_1=0.3$     &                            $0.299$ ($0.069$) & $0.291$ ($0.058$) & $$-$0.001$ & $$-$0.009$ & $0.046$ & $0.033$ & $0.960$ & $1$ \\ [2ex]

 400 & High & $\beta_{0}=-1.185$     &           $$-$1.182$ ($0.653$) & $$-$1.194$ ($0.637$) & $0.003$ & $$-$0.009$ & $0.695$ & $0.655$ & $0.960$ & $0.948$ \\
 &  & $\beta_{1}=1.057$          &                      $1.104$ ($0.341$) & $1.084$ ($0.332$) & $0.047$ & $0.027$ & $0.346$ & $0.253$ & $0.948$ & $0.984$ \\
 &  & $\phi=3$                          &                      $3.215$ ($1.282$) & $3.147$ ($1.224$) & $0.215$ & $0.147$ & $1.283$ & $0.775$ & $0.964$ & $1$ \\
 &  & $\alpha_0=-1.5$              &                      $$-$1.500$ ($0.162$) & $$-$1.500$ ($0.163$) & $0.0004$ & $$-$0.0003$ & $0.166$ & $0.164$ & $0.932$ & $0.932$ \\
 &  & $\alpha_1=0.5$               &                      $0.503$ ($0.062$) & $0.502$ ($0.062$) & $0.003$ & $0.002$ & $0.065$ & $0.058$ & $0.940$ & $0.956$ \\
 &  & $\gamma_1=0.3$           &                       $0.294$ ($0.041$) & $0.295$ ($0.041$) & $$-$0.006$ & $$-$0.005$ & $0.040$ & $0.027$ & $0.932$ & $1$ \\ [2ex]
 
 400 & Low & $\beta_{0}=0.582$     &             $0.614$ ($0.665$) & $0.610$ ($0.665$) & $0.032$ & $0.028$ & $0.673$ & $0.592$ & $0.936$ & $0.948$ \\
 & & $\beta_{1}=1.002$           &                    $0.999$ ($0.265$) & $1.006$ ($0.265$) & $$-$0.003$ & $0.004$ & $0.274$ & $0.224$ & $0.928$ & $0.980$ \\ 
 &  & $\phi=3$                      &                        $3.000$ ($0.828$) & $3.020$ ($0.817$) & $0.0001$ & $0.020$ & $0.810$ & $0.527$ & $0.944$ & $0.992$ \\
 &  & $\alpha_0=-1.5$             &                     $$-$1.486$ ($0.132$) & $$-$1.486$ ($0.133$) & $0.014$ & $0.014$ & $0.131$ & $0.127$ & $0.960$ & $0.972$ \\ 
 &  & $\alpha_1=0.5$               &                    $0.500$ ($0.053$) & $0.496$ ($0.053$) & $$-$0.0003$ & $$-$0.004$ & $0.050$ & $0.047$ & $0.960$ & $0.976$ \\ 
 &  & $\gamma_1=0.3$             &                   $0.301$ ($0.036$) & $0.299$ ($0.036$) & $0.001$ & $$-$0.001$ & $0.035$ & $0.023$ & $0.960$ & $1$ \\ [2ex]
 
  200 & High & $\beta_{0}=-1.182$     &            $$-$1.164$ ($0.747$) & $$-$1.190$ ($0.697$) & $0.019$ & $$-$0.008$ & $0.764$ & $0.677$ & $0.984$ & $0.992$ \\ 
 &  & $\beta_{1}=0.681$           &                      $0.742$ ($0.407$) & $0.721$ ($0.392$) & $0.061$ & $0.040$ & $0.357$ & $0.264$ & $0.956$ & $0.996$ \\ 
 &  & $\phi=1.5$                      &                       $1.755$ ($1.757$) & $1.676$ ($1.646$) & $0.255$ & $0.176$ & $1.266$ & $0.748$ & $0.960$ & $1$ \\
 &  & $\alpha_0=-1.5$             &                       $$-$1.497$ ($0.185$) & $$-$1.494$ ($0.183$) & $0.003$ & $0.006$ & $0.188$ & $0.183$ & $0.936$ & $0.956$ \\
 &  & $\alpha_1=0.5$               &                       $0.500$ ($0.076$) & $0.498$ ($0.076$) & $$-$0.0001$ & $$-$0.002$ & $0.069$ & $0.062$ & $0.964$ & $0.984$ \\
 &  & $\gamma_1=0.3$             &                      $0.289$ ($0.057$) & $0.292$ ($0.055$) & $$-$0.011$ & $$-$0.008$ & $0.046$ & $0.034$ & $0.932$ & $0.992$ \\ [2ex]
 
  400 & High & $\beta_{0}=-1.182$     &            $$-$1.242$ ($0.457$) & $$-$1.231$ ($0.446$) & $$-$0.059$ & $$-$0.048$ & $0.447$ & $0.419$ & $0.984$ & $0.996$ \\
 & & $\beta_{1}=0.681$           &                       $0.741$ ($0.269$) & $0.711$ ($0.264$) & $0.060$ & $0.031$ & $0.281$ & $0.192$ & $0.932$ & $1$ \\
 & & $\phi=1.5$                      &                         $1.615$ ($1.083$) & $1.544$ ($1.064$) & $0.115$ & $0.044$ & $0.941$ & $0.548$ & $0.972$ & $1$ \\
 & & $\alpha_0=-1.5$             &                         $$-$1.500$ ($0.125$) & $$-$1.501$ ($0.125$) & $0.0004$ & $$-$0.001$ & $0.125$ & $0.122$ & $0.956$ & $0.956$ \\ 
 & & $\alpha_1=0.5$               &                        $0.500$ ($0.053$) & $0.501$ ($0.052$) & $$-$0.0003$ & $0.001$ & $0.050$ & $0.042$ & $0.952$ & $0.980$ \\
 & & $\gamma_1=0.3$             &                       $0.293$ ($0.037$) & $0.295$ ($0.037$) & $$-$0.007$ & $$-$0.005$ & $0.036$ & $0.026$ & $0.924$ & $0.988$ \\ [2ex]
 
\hline			
\end{tabular*}
\label{table:T8}\\
\end{sidewaystable}

\begin{table}  [!htbp] 
\color{blue}
\caption{\textcolor{blue}{Comparison of CPU time (in seconds) between SEM and MCEM algorithms}}
\begin{tabular*}{\textwidth}{@{\extracolsep{\fill}} ccccc}\\
\hline                                 
$n$ & Cure Rate & $\phi$     & \multicolumn{2}{c}{CPU time (in seconds)} \\ \cline{4-5} 
& & & \multicolumn{1}{c}{SEM}&\multicolumn{1}{c}{MCEM} \\     
\hline                       
200 & High & 3 & 10.563 & 54.261 \\
400 & High & 3 & 14.734 & 84.355 \\
400 & Low & 3 & 11.886 &  55.994 \\
200 & High & 1.5 & 10.742 & 63.533 \\
400 & High & 1.5 & 15.174 & 96.670 \\
\hline
\end{tabular*}
\label{table:T9}
\end{table}

\section{Application: breast cancer data}

\begin{figure}[ht!]
\centering
\includegraphics[width=0.8\linewidth]{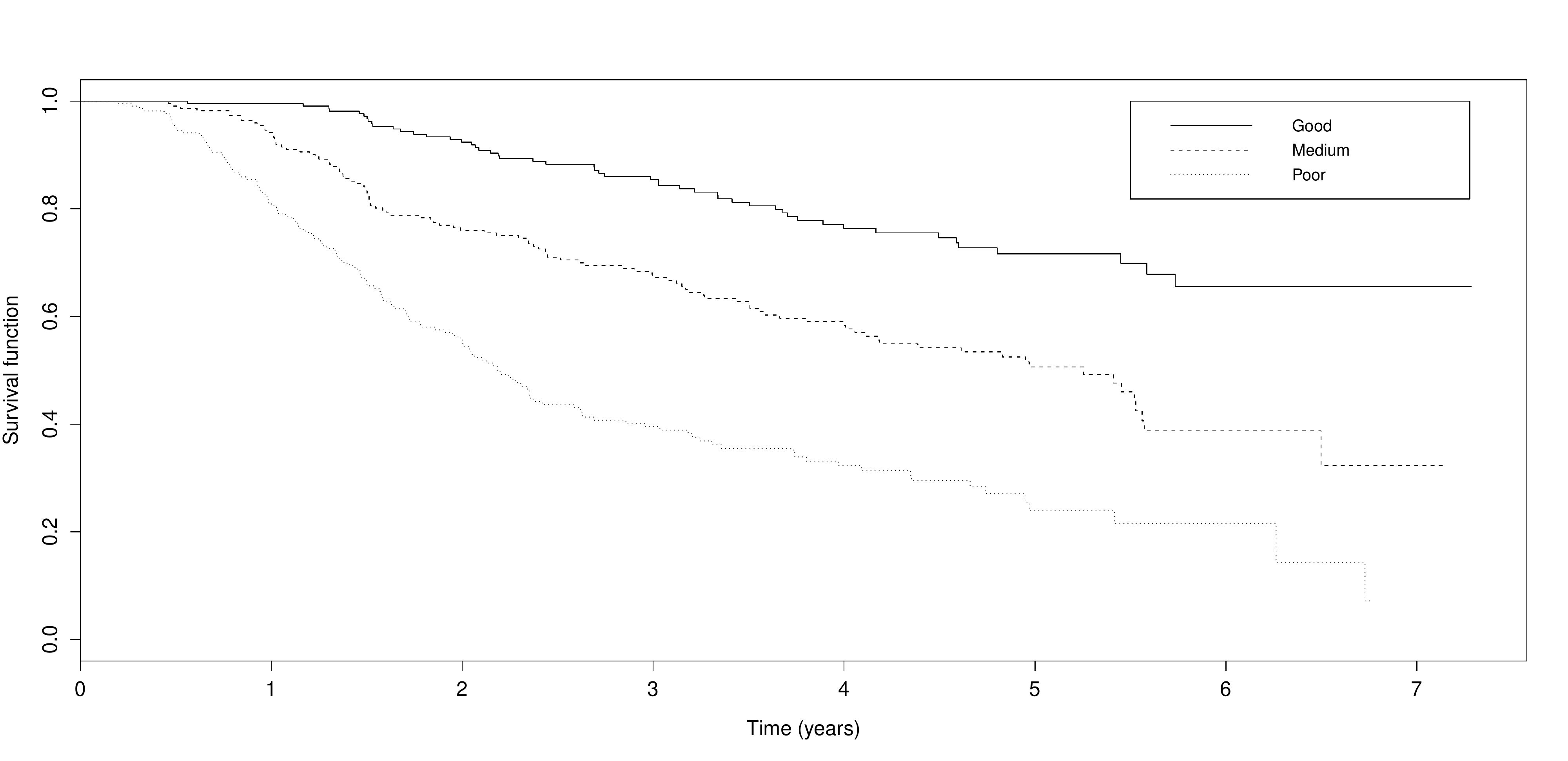}
\caption{Kaplan-Meier plot of survival curves stratified by different prognostic groups}
\label{figure:KM}
\end{figure}

In this section, we apply our proposed SEM algorithm to a real data on breast cancer readily available in R package ``flexsurv''.$^{43}$ The dataset represents the survival times, defined as the time to death or the censoring time, of 686 patients with primary node positive breast cancer. We consider the variable prognostic group ($x=$ 1 representing ``good'', $x=$ 2 representing ``medium'', and $x=$ 3 representing ``poor'') as the covariate in our application. The observed time has a mean of 3.08 years and a standard deviation of 1.76 years. The percentage of censored observations is 56\%. In Figure \ref{figure:KM}, we present the Kaplan-Meier curves of the survival function stratified by prognostic group. It can be seen that the survival curves do not intersect and there is a clear trend in the survival of patients belonging to different prognostic groups. To be specific, we can see that patients in better group category have higher survival times. Furthermore, the survival curves, more specifically for groups ``good'' and ``medium'', level off to non-zero proportions, indicating the presence of cure rates for these groups.

To calculate the initial values of the parameters associated with the cure rates, i.e., $\beta_0,$ $\beta_1$ and $\phi$, we equate the non-parametric estimates of cure rates for the three groups (obtained from the Kaplan-Meier curves in Figure \ref{figure:KM}) to their corresponding theoretical expressions. Then, we solve the three equations to find the values of $\beta_0,$ $\beta_1$ and $\phi$, which are taken as their initial values. To find the initial values of the parameters associated with the lifetime distribution, i.e., $\gamma_1$ and $\gamma_2$, we first equate the theoretical expressions of the mean and variance corresponding to the Weibull density function in \eqref{ft} to the observed mean and variance of the breast cancer survival times. Then, we solve these two equations to find the values of $\gamma_1$ and $\gamma_2$. The value of $\gamma_1$ that we obtain is taken as the initial value of the parameter $\gamma_1$. Note that $\gamma_2$ is linked with the covariate through $\gamma_2 = e^{\alpha_0 + \alpha_1 x}$. As such, to find the initial values of $\alpha_0$ and $\alpha_1$, we have to perform additional steps. For this purpose, we consider any two of the three groups, say, we consider groups 1 and 3. Then, we equate the mean observed survival times of these two groups to their corresponding theoretical means. On solving these two equations, we finally get the initial values of $\alpha_0$ and $\alpha_1$.

In Table \ref{table:mle}, we present the SEM estimates of the model parameters. For the purpose of comparison, we also present the EM estimates. To get the EM estimates, we use a profile likelihood set of $\phi$ as $\{0.1,0.2,\cdots,10\}$. To get the SEM estimates, after a preliminary study, we use 10000 iterations and the first 6000 iterations are considered as burn-in. From Table \ref{table:mle}, we note that the parameter $\beta_1$ is significant (from the 95\% confidence interval of $\beta_1$) and its estimate is positive. This clearly implies that the mean number of competing risks increase with a decrease in the prognostic group status, which is what we would expect. From the 95\% confidence interval of the parameter $\alpha_1$, we can conclude that the distribution of the progression times is homogeneous for all three group categories. Using the SEM and EM estimates, the maximized log-likelihood value turns out to be -790.690 and -790.989, respectively. Thus, the proposed SEM algorithm performs better than the EM algorithm for the considered breast cancer data. In Table \ref{table:cure}, we present the estimation results corresponding to the cure rates. As already seen in Figure \ref{figure:KM}, Table \ref{table:cure} confirms that the cure rates for the groups ``good'' and ``medium" are significantly larger than that for the group ``poor". Note that the asymptotic confidence intervals corresponding to $p_{01}$ and $p_{02}$ are non-overlapping, indicating that the cure rate of patients belonging to the group ``good" is significantly different (higher) from the cure rate of patients belonging to the group ``medium".

\begin{figure}[ht!]
\centering
\includegraphics[width=0.8\linewidth]{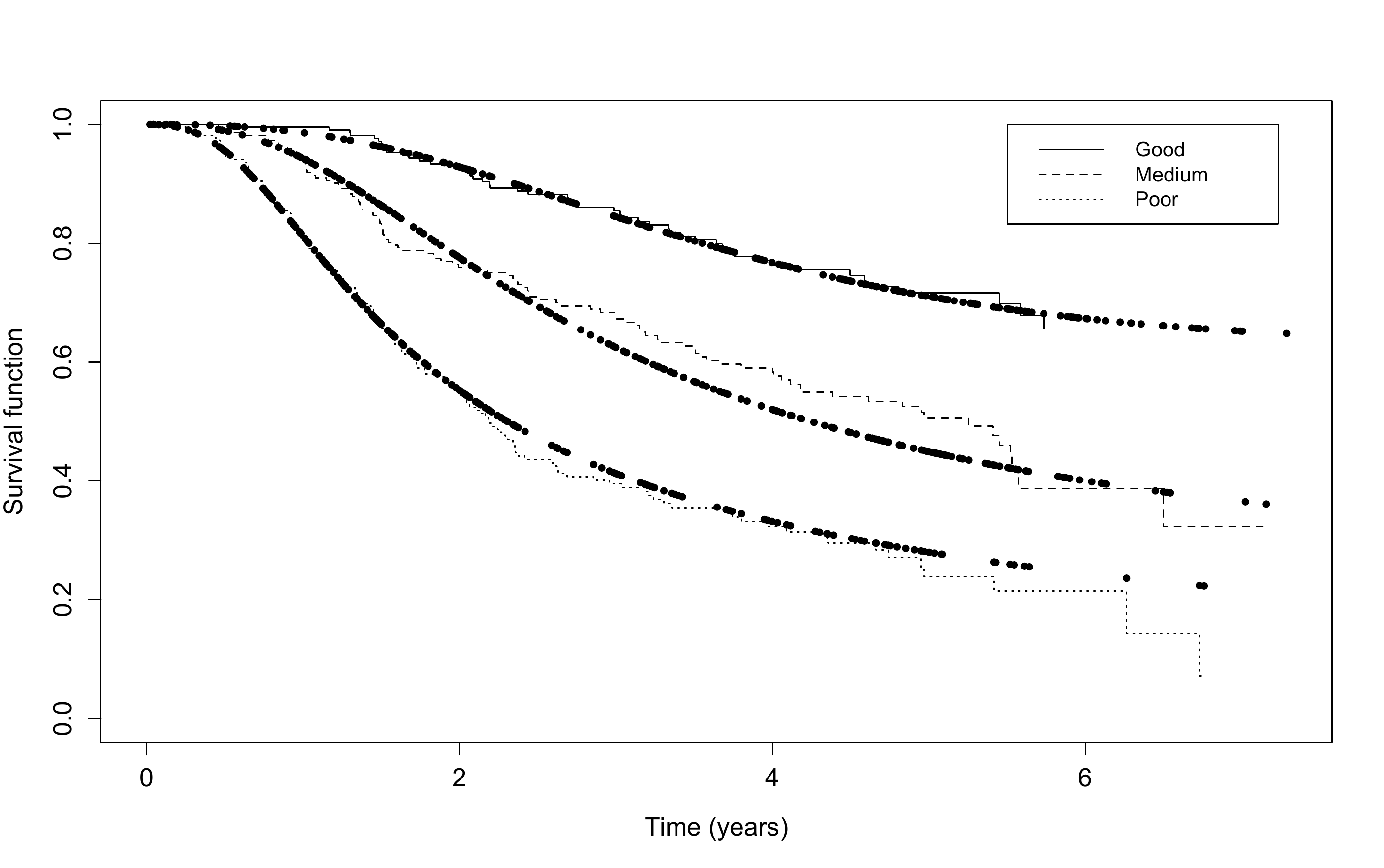}
\caption{Estimated survival curves superimposed on non-parametric Kaplan-Meier survival curves corresponding to the SEM estimates}
\label{figure:surv}
\end{figure}

\begin{figure}[ht!]
\centering
\includegraphics[width=0.8\linewidth]{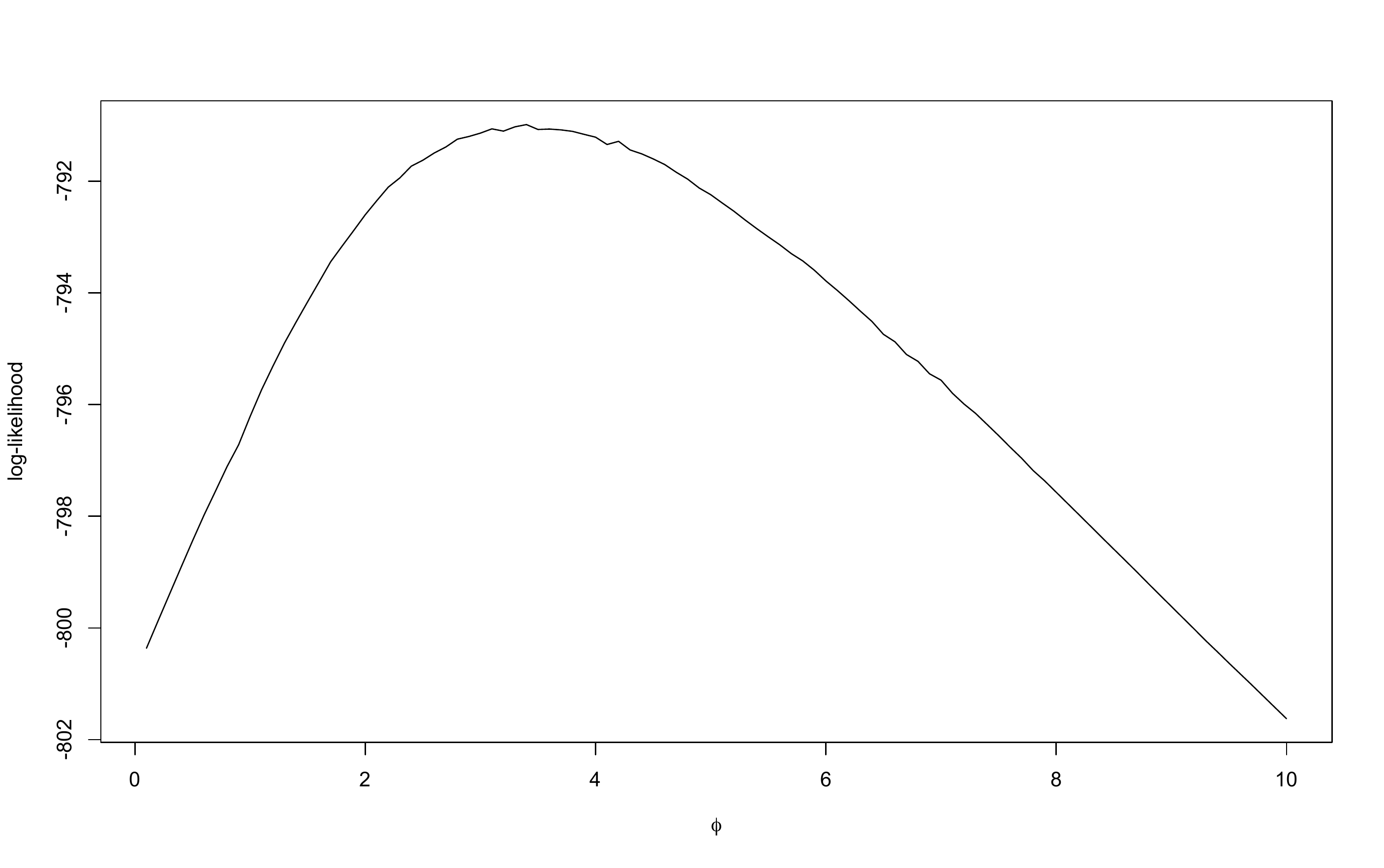}
\caption{Profile likelihood plot for the parameter $\phi$}
\label{figure:prof}
\end{figure}

\begin{figure}[ht!]
\centering
\includegraphics[width=0.8\linewidth]{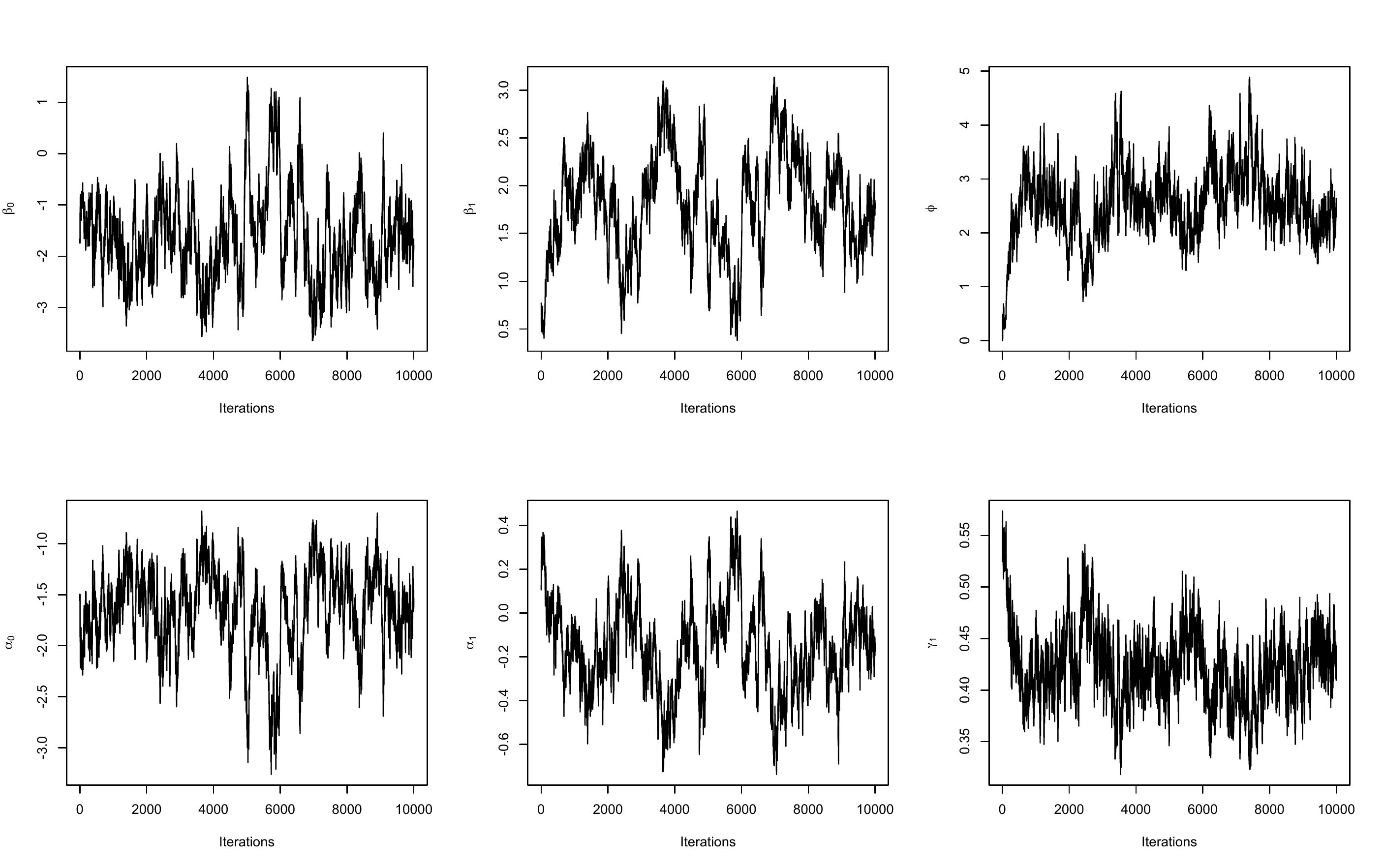}
\caption{Parameter evolutions in the SEM algorithm}
\label{figure:iter}
\end{figure}

In Figure \ref{figure:surv}, we present the plot of estimated survival curves (using the SEM estimates) stratified by prognostic groups and superimpose them on the Kaplan-Meier curves. It is clear that the parametric survival curves show a close concordance with the non-parametric Kaplan-Meier curves for each of the three prognostic groups. Figure \ref{figure:prof} shows the profile likelihood plot of $\phi$. Figure \ref{figure:iter} shows the evolution paths of the parameters in the SEM algorithm for the negative binomial cure rate model. We note that the SEM iterations oscillate without any indication of any significant upward or downward trend.$^{36}$

\begin{figure}[ht!]
\centering
\includegraphics[width=0.8\linewidth]{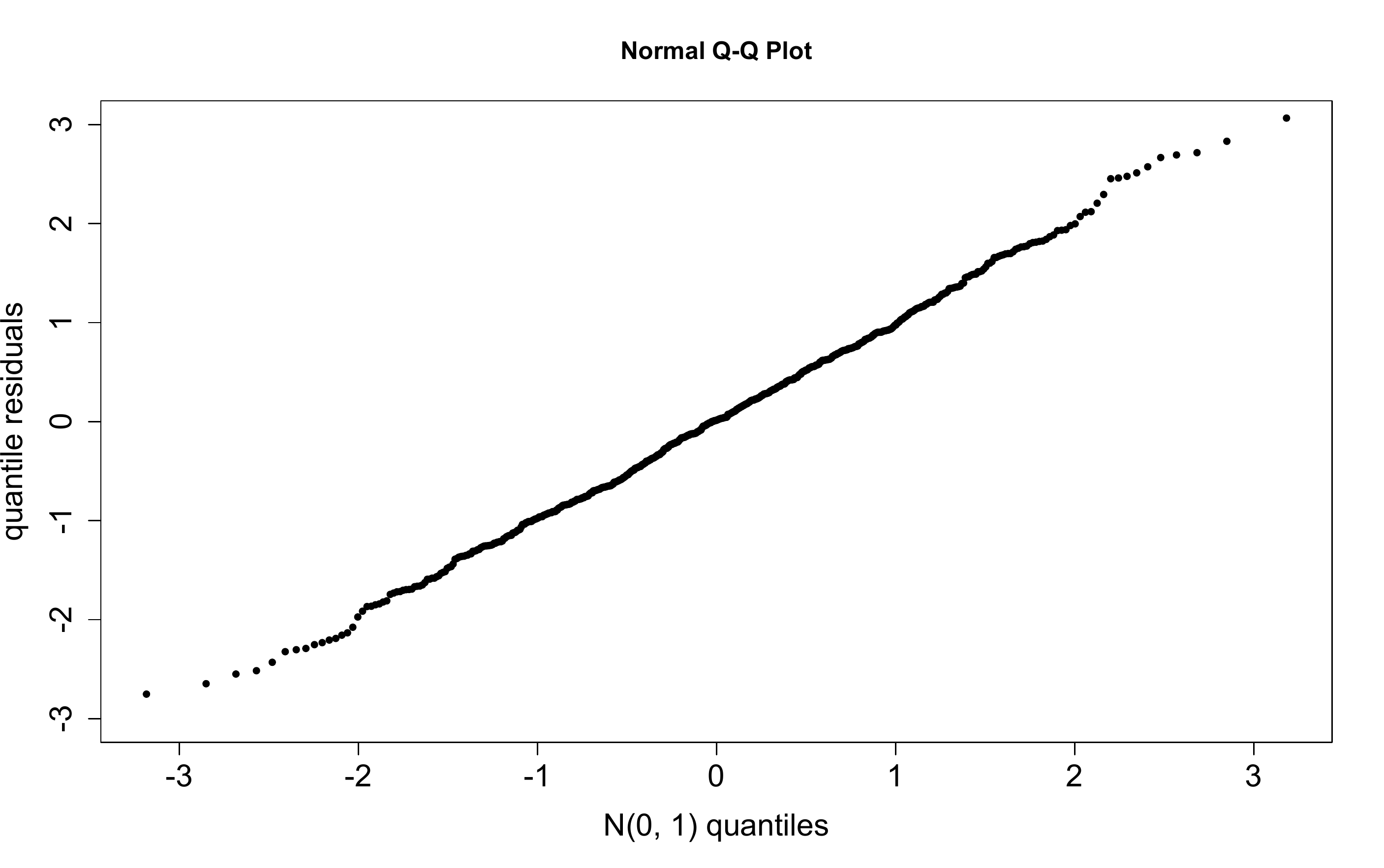}
\caption{QQ plot of the normalized randomized quantile residuals corresponding to the SEM estimates}
\label{figure:qq}
\end{figure}

We also check for the goodness-of-fit of our model. This is done by using the calculated normalized randomized quantile residuals.$^{44}$ The quantile-quantile plot is presented in Figure \ref{figure:qq} and it is clear that the negative binomial cure rate model with Weibull lifetimes provide a very good fit to the breast cancer data. In Figure \ref{figure:qq}, each point corresponds to the median of five sets of ordered residuals. Finally, we test for the normality of residuals using the Kolmogorov-Smirnov test and the p-value turns out to be 0.959, suggesting very strong evidence for the normality of residuals.
\begin{table}  [!htbp] 
\caption{Estimates, standard errors and 95$\%$ confidence intervals (CI) of model parameters for the breast cancer data}
\begin{tabular*}{\textwidth}{@{\extracolsep{\fill}} l l l l l l l}\\
\hline                                 
Parameter                         & \multicolumn{2}{c}{Estimates} & \multicolumn{2}{c}{Standard errors} & \multicolumn{2}{c}{95\% CI} \\  \cline{2-3} \cline{4-5} \cline{6-7}
& \multicolumn{1}{c}{SEM}&\multicolumn{1}{c}{EM} & \multicolumn{1}{c}{SEM}&\multicolumn{1}{c}{EM} & \multicolumn{1}{c}{SEM}&\multicolumn{1}{c}{EM} \\ 
\hline                                 
$\beta_{0}$                  & -2.756 &  -2.346      &  1.194 &  1.115      &  (-5.096, -0.415) &  (-4.532, -0.161) \\
$\beta_{1}$                  & 2.801  & 2.510     &  1.097 &   0.866        & (0.650, 4.951) &   (0.812, 4.207) \\
$\phi$                           & 3.281  &  3.400   &  0.939 &  0.963         & (1.440, 5.121) &   (1.511, 5.288) \\
$\alpha_0$                   & -1.152 & -1.334     &  0.465 & 0.447       & (-2.063, -0.241) &  (-2.211, -0.458)  \\
$\alpha_1$                    & -0.488 &  -0.357    &  0.361 &  0.282    & (-1.196, 0.220) &    (-0.910, 0.196) \\
$\gamma_1$                &   0.381 &  0.377   & 0.047 & 0.046         & (0.290, 0.473)  &    (0.287, 0.468)     \\
\hline
\end{tabular*}
\label{table:mle}
\end{table}

\begin{table}  [!htbp] 
\caption{Estimates, standard errors and 95$\%$ confidence intervals (CI) of cure rates for patients belonging to different prognostic groups}
\begin{tabular*}{\textwidth}{@{\extracolsep{\fill}} l l l l l l l l}\\
\hline                                 
Group & Cure rate        &  \multicolumn{2}{c}{Estimates} & \multicolumn{2}{c}{Standard errors} & \multicolumn{2}{c}{95\% CI} \\  \cline{3-4} \cline{5-6} \cline{7-8}
& & \multicolumn{1}{c}{SEM}&\multicolumn{1}{c}{EM} & \multicolumn{1}{c}{SEM}&\multicolumn{1}{c}{EM} & \multicolumn{1}{c}{SEM}&\multicolumn{1}{c}{EM} \\ 
\hline                                 
Good & $p_{01}$              & 0.635 &   0.623    &  0.067  &   0.078    &  (0.504, 0.767) & (0.469, 0.776)   \\
Medium & $p_{02}$          & 0.291  &  0.316    &  0.071 &    0.058   & (0.151, 0.431)  & (0.203, 0.429)    \\
Poor & $p_{03}$              &  0.124  &  0.152    &  0.062  &   0.056   & (0.003, 0.246) &  (0.042, 0.261)  \\
\hline
\end{tabular*}
\label{table:cure}
\end{table}

\section{Conclusion and future work}

In this paper, we develop a new estimation method for cure rate model with latent competing risks. Motivated by a recent work,$^{27	}$ we consider the unobserved competing risks as the missing data but develop a variation of the EM algorithm, called the stochastic EM (SEM) algorithm, to calculate the MLEs of the model parameters. Assuming the number of competing risks to follow a negative binomial distribution, we show that the SEM algorithm avoids calculation of complicated expectations. This can be looked as a major advantage of the SEM algorithm over the well-known EM algorithm. Through a Monte Carlo simulation study, we show that the SEM algorithm can retrieve the true parameter values quite accurately with small bias. The standard error and the RMSE of the estimators of the model parameters both decrease with an increase in sample size. Furthermore, the coverage probabilities are close to the nominal level. In this context, the EM algorithm results in over-coverage corresponding to the parameter $\phi$. The over-coverage is also noticed for other model parameters when the true cure rates are high. Overall, the proposed SEM algorithm can be considered as the preferred algorithm. Through the real breast cancer data, we show that the performance of the SEM algorithm is better when compared to the EM algorithm. The use of SEM algorithm in the context of cure rate models is new$^{28}$ and we believe that this work will motivate researchers to consider SEM algorithm for more complicated cure rate models. For instance, one can think of developing the SEM algorithm and the associated likelihood inference for cure rate models with destruction of competing risks.$^{15}$ Furthermore, one can also consider cure rate models with interval censored data and develop the inferential framework based on the SEM algorithm. We are currently working on these and hope to report the findings in a future paper.

\color{blue}

\section*{Data availability statement}

The breast cancer data is available in R package ``flexsurv''. The computational codes are available in the supplementary material of this article.

\section*{Acknowledgement}

The author would like to thank the Associate Editor and two anonymous reviewer's for their constructive comments and feedback on an earlier version of this manuscript.

\color{black}

\section*{Conflict of interest}

The author has declared no conflict of interest.

\section*{References}

\begin{enumerate}

\item Boag JW. Maximum likelihood estimates of the proportion of patients cured by cancer therapy. {\it J R Stat Soc Series B Stat Methodol.} 1949;{11}:15$-$53.
\item Berkson J, Gage RP. Survival curve for cancer patients following treatment. {\it  J Am Stat Assoc.} 1952;{47}:501$-$515.
\item Chen MH, Ibrahim JG, Sinha D. A new Bayesian model for survival data with a surviving fraction. {\it  J Am Stat Assoc.} 1999;{94}:909$-$919.
\item Kannan N, Kundu D, Nair P, Tripathi, RC. The generalized exponential cure rate model with covariates. {\it J Appl Stat.} 2010;{37}:1625$-$1636.
\item Sy JP, Taylor JMG. Estimation of a Cox proportional hazards cure model. {\it Biometrics.} 2000;{56}:227$-$236.
\item Rodrigues J, Castro M, Cancho VG, Balakrishnan N. COM-Poisson cure rate survival models and an application to cutaneous melanoma data. {\it J Stat Plan Inference.} 2009;{139}:3605$-$3611.
\item Yin G, Ibrahim JG. Cure rate models: a unified approach. {\it Can J Stat.} 2005;{33}:559$-$570.
\item Balakrishnan N, Pal S. Lognormal lifetimes and likelihood-based inference for flexible cure rate models based on COM-Poisson family. {\it Comput Stat Data Anal.} 2013;{67}:41$-$67.
\item Balakrishnan N, Pal, S. Expectation maximization-based likelihood inference for flexible cure rate models with Weibull lifetimes. {\it Stat Methods Med Res.} 2016;{25}:1535$-$1563.
\item Kuk AYC, Chen CH. A mixture model combining logistic regression with proportional hazards regression. {\it Biometrika.} 1992;{79}:531$-$541.
\item Li CS, Taylor JMG. A semi-parametric accelerated failure time cure model. {\it Stat Med.} 2002;{21}:3235$-$3247.
\item Balakrishnan N, Barui S, Milienos FS. Proportional hazards under Conway–Maxwell-Poisson cure rate model and associated inference. {\it Stat Methods Med Res.} 2017;{26}:2055$-$2077.
\item Balakrishnan N, Koutras MV, Milienos FS, Pal S. Piecewise linear approximations for cure rate models and associated inferential issues. {\it Methodol Comput Appl Probab.} 2016;{18}:937$-$966.
\item Klebanov LB, Rachev ST, Yakovlev AY. A stochastic model of radiation carcinogenesis: latent time distributions and their properties. {\it Math Biosci.} 1993;{113}:51$-$75.
\item Rodrigues J, Castro M, Balakrishnan N, Cancho VG. Destructive weighted Poisson cure rate models. {\it Lifetime Data Anal.} 2011;{17}:333$-$346.
\item Pal S, Majakwara J, Balakrishnan N. An EM algorithm for the destructive COM-Poisson regression cure rate model. {\it Metrika.} 2018;{81}:143$-$171.
\item Pal S, Balakrishnan N. Destructive negative binomial cure rate model and EM-based likelihood inference under Weibull lifetime. {\it Stat Probab Lett.} 2016;{116}:9$-$20.
\item Pal S, Balakrishnan N. Likelihood inference for the destructive exponentially weighted Poisson cure rate model with Weibull lifetime and an application to melanoma data. {\it Comput Stat.} 2017;{ 32}:429$-$449.
\item Pal S, Balakrishnan N. Expectation maximization algorithm for Box-Cox transformation cure rate model and assessment of model mis-specification under Weibull lifetimes. {\it IEEE J Biomed Health Inform.} 2018;{22}:926$-$934.
\item Majakwara J, Pal S. On some inferential issues for the destructive COM-Poisson-generalized gamma regression cure rate model. {\it Commun Stat Simul Comput.} 2019;{48}:3118$-$3142.
\item Rodrigues J, Cancho VG, Castro M, Balakrishnan N. A Bayesian destructive weighted Poisson cure rate model and an application to a cutaneous melanoma data. {\it Stat Methods Med Res.} 2012;{21}:585$-$597.
\item Cancho VG, Louzada F, Ortega EM. The power series cure rate model: an application to a cutaneous melanoma data. {\it Commun Stat Simul Comput.} 2013;{42}:586$-$602.
\item Gallardo DI, Bolfarine H, Pedroso-de Lima AC. An EM algorithm for estimating the destructive weighted Poisson cure rate model. {\it J Stat Comput Simul.} 2016;{86}:1497$-$1515.
\item Pal S, Balakrishnan N. Likelihood inference for COM-Poisson cure rate model with interval-censored data and Weibull lifetimes. {\it Stat Methods Med Res.} 2017;{26}:2093$-$2113.
\item Wiangnak P, Pal S. Gamma lifetimes and associated inference for interval censored cure rate model with COM-Poisson competing cause. {\it Commun Stat Theory Methods.} 2018;{47}:1491$-$1509.
\item Balakrishnan N, Pal S. EM algorithm-based likelihood estimation for some cure rate models. {\it J Stat Theory Pract.} 2012;{6}:698$-$724.
\item Gallardo DI, Romeo JS, Meyer R. A simplified estimation procedure based on the EM algorithm for the power series cure rate model. {\it Commun Stat Simul Comput.} 2017;{46}:6342$-$6359.
\item Davies K, Pal S, Siddiqua, JA. Stochastic EM algorithm for generalized exponential cure rate model and an empirical study. {\it J Appl Stat.} 2020. https://doi.org/10.1080/02664763.2020.1786676. Accessed June 15, 2021.
\item Ortega EMM, Barriga GDC, Hashimoto EM, Cancho VG, Cordeiro GM. A new class of survival regression models with cure fraction. {\it J Data Sci.} 2014;{12}:107$-$136.
\item McLachlan GJ, Krishnan T. {\it The EM Algorithm and Extensions.} 2nd ed. Hoboken, NJ: John Wiley \& Sons; 2008.
\item Celeux G, Diebolt J. The SEM algorithm: a probabilistic teacher algorithm derived from the EM algorithm for the mixture problem. {\it Comput Stat.} 1985;{2}:73$-$82.
\item Bordes L, Chauveau D, Vandekerkhove, P. A stochastic EM algorithm for a semiparametric mixture model. {\it Comput Stat Data Anal.} 2007;{51}:5429$-$5443.
\item Cariou C, Chehdi K. Unsupervised texture segmentation/classification using 2-d autoregressive modeling and the stochastic expectation–maximization algorithm. {\it Pattern Recognit Lett.} 2008;{29}:905$-$917.
\item Diebolt J, Celeux G. Asymptotic properties of a stochastic EM algorithm for estimating mixing proportions. {\it Stoch Model.} 1993;{9}:599$-$613.
\item Chauveau D. A stochastic EM algorithm for mixtures with censored data. {\it J Stat Plan Inference.} 1995;{46}:1$-$25.
\item Nielsen SF. The stochastic EM algorithm: Estimation and asymptotic results. {\it Bernoulli.} 2000;{6}:457$-$489.
\item Svensson I, Sjostedt-deLuna S. Asymptotic properties of a stochastic EM algorithm for mixtures with censored data. {\it J Stat Plan Inference.} 2010;{140}:111$-$127.
\item Diebolt J, Ip E. {\it Markov Chain Monte Carlo in practice}. New York: Springer; 1996.
\item Yang Y, Ng HKT, Balakrishnan N. A stochastic expectation-maximization algorithm for the analysis of system lifetime data with known signature. {\it Comput Stat.} 2016;{31}:609$-$641.
\item Marschner IC. On stochastic versions of the EM algorithm. {\it Biometrika.} 2001;{88}:281$-$286.
\item Ye Z, Ng HKT. On analysis of incomplete field failure data. {\it Ann Appl Stat.} 2014;{8}:1713$-$1727.
\item Gallardo DI, Gomez YM, Castro M. A note on the EM algorithm for estimation in the destructive negative binomial cure rate model. {\it J Stat Comput Simul.} 2017;{87}:2291$-$2297.	
\item Sauerbrei W, Royston P. Building multivariable prognostic and diagnostic models: transformation of the predictors using fractional polynomials. {\it J R Stat Soc Ser A Stat Soc.} 1999;{162}:71$-$94.
\item Dunn PK, Smyth GK. Randomized quantile residuals. {\it J Comput Graph Stat.} 1996;{5}:236$-$244.

\end{enumerate}

\end{document}